\documentclass{article} % For LaTeX2e
\usepackage{iclr2025_conference,times}

\usepackage{amsmath,amsfonts,bm}

\def\eqref#1{equation~\ref{#1}}
\def\1{\bm{1}}

\DeclareMathAlphabet{\mathsfit}{\encodingdefault}{\sfdefault}{m}{sl}
\SetMathAlphabet{\mathsfit}{bold}{\encodingdefault}{\sfdefault}{bx}{n}

\usepackage{xcolor}
\usepackage[pagebackref,breaklinks,colorlinks,citecolor=blue]{hyperref}
\usepackage{url}
\usepackage[T1]{fontenc}
\usepackage{array,graphicx}
\usepackage{booktabs}
\usepackage{colortbl}
\usepackage{subfig}
\usepackage{pifont}

\newcommand*\rot{\rotatebox{90}}
\newcommand*\OK{\ding{51}}

\usepackage[most]{tcolorbox}
\tcbuselibrary{skins,breakable}
\newtcolorbox{mybox}[2][]{breakable,sharp corners, skin=enhancedmiddle jigsaw,parbox=false,
boxrule=0mm,leftrule=2mm,boxsep=0mm,arc=0mm,outer arc=0mm,attach title to upper,
after title={.\ }, coltitle=black,colback=gray!10,colframe=black, title={#2},
fonttitle=\bfseries,#1}

\title{Data-Centric AI Governance: Addressing the Limitations of Model-Focused Policies}

\author{Ritwik Gupta \enspace Leah Walker \enspace Rodolfo Corona \enspace Stephanie Fu \enspace Suzanne Petryk\\
\vspace{-0.8em}\\
{\bf Janet Napolitano \enspace Trevor Darrell \enspace Andrew W. Reddie}\\
\vspace{-0.4em}\\
University of California, Berkeley}

\iclrfinalcopy % Uncomment for camera-ready version, but NOT for submission.
\arxiv
\begin{document}

\maketitle

\begin{abstract}
Current regulations on powerful AI capabilities are narrowly focused on ``foundation'' or ``frontier'' models. However, these terms are vague and inconsistently defined, leading to an unstable foundation for governance efforts. Critically, policy debates often fail to consider the data used with these models, despite the clear link between data and model performance. Even (relatively) ``small'' models that fall outside the typical definitions of foundation and frontier models can achieve equivalent outcomes when exposed to sufficiently specific datasets. In this work, we illustrate the importance of considering dataset size and content as essential factors in assessing the risks posed by models both today and in the future. More broadly, we emphasize the risk posed by over-regulating reactively and provide a path towards careful, quantitative evaluation of capabilities that can lead to a simplified regulatory environment.
\end{abstract}

\section{The Shortcomings of Today’s AI Governance}
The past decade has seen a rapid burst of commercial AI products, such as Google Translate and OpenAI’s ChatGPT, delivering new capabilities into the hands of the public. As AI has made its way to wider audiences, it has continued its rapid pace of development, giving everyday users what were previously highly specialized computing tools and capabilities. This raises questions for governments, academics, and commercial labs about whether certain AI capabilities or behaviors should be deemed as too ``risky'' for public access~\citep{draganFrontierSafetyFramework2024}.

Today’s AI governance efforts have coalesced around the terms ``frontier'', ``foundation'', ``dual-use'', and ``general purpose'' to describe the largest and most capable of these models. In policy papers and legislation, models described by these terms are subject to additional scrutiny and regulatory interest. These terms are usually synonymous with the most cutting-edge models of today including OpenAI’s ChatGPT~\citep{brownLanguageModelsAre2020, openaiGPT4TechnicalReport2024}, Meta’s LLaMA~\citep{touvronLLaMAOpenEfficient2023}, and Google’s Gemini~\citep{geminiteamGeminiFamilyHighly2024}. Although there is general agreement for the types of AI-accelerated risks that regulations aim to curtail, there is much less clarity and consensus in concrete definitions for such models. In an effort to define the characteristics of these large, capable models, a number of policy documents have focused on parameter counts and/or FLOPs, measures of model size and compute requirement~\citep{europeanunionRegulationEU20242024}.

We argue that this approach is short-sighted for three reasons. First, there is no consistent definition of ``frontier'', ``foundation'', ``dual-use'', and ``general purpose'' models with regards to FLOPs or parameter count. As discussed below in Section \ref{sec:definition-unstable}, this lack of definitional clarity has led to a regulatory and governance landscape with varying ceilings for what constitutes a covered capability. Second, as machine learning advances, models are becoming more efficient, requiring fewer parameters and FLOPs to achieve the same tasks. This means that the next generation of models could be more capable while falling below regulatory ceilings. Finally, the focus on the largest and most compute-intensive models ignores the fact that smaller models can be just as capable in niche, and potentially risky, areas as their larger counterparts. These factors culminate in inadvertent loopholes that powerful capabilities can slip through, rendering expensive regulatory efforts not only useless but potentially detractory from beneficial uses of AI technologies.

Analyzing AI purely as a function of models is an outdated view. The broader field of machine learning has recognized the role of data as a direct indicator of model performance~\citep{hoffmannEmpiricalAnalysisComputeoptimal2022, ngDataCentricAICompetition2021}, suggesting that dataset quality and size should also be included as factors in conversations surrounding model capabilities.

In this paper, we first discuss the limitations of the current model-focused governance ecosystem. Then, we demonstrate the value of a data-focused approach to AI governance. In particular, we present experiments corroborating the role that dataset size plays in model capability. Finally, we propose legal and technical approaches to AI governance rooted in our understanding of the dataset-model capability relationship.

\section{Definitional Challenges and Flawed Limits in AI Governance}
Over time, the capabilities of digital systems have improved in tandem with hardware advancements~\citep{hilbertWorldsTechnologicalCapacity2011}. This is also true of machine learning models; simple two-parameter logistic regressions have evolved to deep neural networks with trillions of parameters, enabled by the advent of graphics processing units (GPUs) and high-speed memory\footnote{This natural progression in size and complexity was formalized as a computing law by Gordon Moore and is colloquially known as Moore’s Law.}~\citep{mooreCrammingMoreComponents1998}. Yet, Moore’s Law has weakened significantly. We have observed periods of stagnation, and even reversal, in aggregate computing trends despite overall progress in terms of effectiveness of outcomes~\citep{leisersonTheresPlentyRoom2020}. The trend of plateauing is also observed in machine learning, and policies aiming to regulate machine learning models solely as a function of continuous growth are flawed, as demonstrated in Section \ref{sec:cape-correlated}.

In addition, much of the conversation around AI regulation has centered itself around the prevention of behaviors that are deemed to be ``harmful'' or otherwise detrimental to society~\citep{dafoeAIGovernanceResearch2018, hoffmanAddingStructureAI2023}.  The mention of ``harm'' is too often unqualified and does not address the capabilities of existing technologies that may already be capable of much of the malicious behavior discussed in AI policy circles today. For example, AI for biological agent design is widely cited as a potential harm~\citep{callawayCouldAIdesignedProteins2024}, yet computational drug discovery has been the norm since the 1980s and has enabled the discovery of drugs such as ritonavir, a medication critical in treating both HIV and COVID-19~\citep{vandrieComputeraidedDrugDesign2007}. The conversation surrounding the use of AI to further societal harms must contextualize the additional marginal risk posed by these methods when compared to existing technologies such as search engines or statistical inference algorithms.

The AI ecosystem’s difficulty in defining and identifying harm extends into inconsistent efforts to identify ``harmful'' or ``risky'' models and regulate them. In the following sections, we demonstrate the shortcomings and inconsistencies of these model-focused AI governance efforts, while identifying key drivers of AI risk that are currently overlooked in modern AI policy.

\subsection{An Unstable Definition Foundation}
\label{sec:definition-unstable}
The use of the terms ``foundation'', ``frontier'', ``dual-use'', and ``general purpose'' to describe machine learning models has arisen in the past few years in an effort to isolate classes of models seen as posing the greatest risk of harm to public safety. In 2021, Stanford University researchers introduced the concept of ``foundation models'' in ``On the Opportunities and Risks of Foundation Models''~\citep{bommasaniOpportunitiesRisksFoundation2022}. The paper uses the term to describe machine learning models trained using self-supervised learning methods on large sets of data to the point that they demonstrate emergent behaviors during inference.

\begin{mybox}{Self-supervised learning}
A model-training technique using unlabeled data rather than relying on external human-provided labels. Oftentimes, labels are generated from the data itself or are inherent in the training process.
\end{mybox}
\begin{mybox}{Inference}
The stage of machine learning in which a model makes predictions on new inputs based on its knowledge up to that point. Model weights are not updated at inference time. 
\end{mybox}

The term ``foundation model'' has spread swiftly throughout the AI research community to a point of saturation where any model trained on a subjectively large set of data can be termed ``foundational.'' More recently, the terms ``frontier model,'' introduced in ``Frontier AI Regulation: Managing Emerging Risks to Public Safety''~\citep{anderljungFrontierAIRegulation2023} and ``dual-use model,'' found in the ``Executive Order on Safe, Secure, and Trustworthy Development and Use of Artificial Intelligence''~\citep{thewhitehouseExecutiveOrderSafe2023} and the EU AI Act~\citep{europeanunionRegulationEU20242024}, have arisen as similar descriptions of large, cutting-edge models with an increased potential for harm. The cross-cutting conclusion of the literature has been that these types of models can pose serious risks to the general public and should be governed as such.

However, despite launching regulatory processes to achieve similar outcomes to curtail risk, various across reports, legislation, and other forms of governance documents lack agreement in definitions utilized to circumscribe powerful AI capabilities. In Table \ref{tab:definitions}, we highlight various impactful papers and policies that have shaped international AI governance. In particular, we highlight the inconsistencies between how influential works which first introduced various terms and thresholds disagree from their actualization in policy proposals.

\begin{table}
\centering
\renewcommand{\arraystretch}{1.5} % Adjust row height
    \begin{tabular}{@{} cl*{5}c @{}}
        & & \rot{Terms} & \rot{SSL} & \rot{\shortstack[l]{Large\\data}}
        & \rot{FLOPs} & \rot{Params.} \\
        \cmidrule{2-7}
        & \citet{bommasaniOpportunitiesRisksFoundation2022} & Foundation                & \OK & \OK &    --      & --\\
        & \citet{anderljungFrontierAIRegulation2023}        & Foundation, Frontier      & \OK & \OK & $>10^{26}$ & --\\
 \rot{\rlap{~\;Terms}}
        & \citet{alstottPreparingFederalResponse2023}       & Frontier                  &  -- &  -- & $>10^{26}$ & --\\\midrule
        & \citet{thewhitehouseExecutiveOrderSafe2023}       & Foundation, ``Dual-Use''\footnotemark[2]  & \OK & \OK & $>10^{26}$ & $>10$B\\
        & \citet{romneyAILetter2024}                        & Frontier, General Purpose &  -- &  -- & $>10^{26}$ & --\\
        & \citet{europeanunionRegulationEU20242024}         & General Purpose           & \OK & \OK & $>10^{25}$ & $>1$B\\
\rot{\rlap{~Governance}}
        & \citet{wienerSB1047SafeSecure2024}                & Frontier                  &  -- &  -- & $>10^{25}$ / $10^{26}$  & --\\
        \cmidrule[1pt]{2-7}
    \end{tabular}
    \caption{Variance in model definitions across policy documents.}
    \label{tab:definitions}
\end{table}
\footnotetext[2]{Despite using the words ``dual-use'', the definition provided in the document are more aligned with accepted definitions of ``general purpose.''}
\setcounter{footnote}{2}

Terminology such as ``foundation'' and ``frontier'' are terms of art that have non-static and contentious definitions, suggesting that utility-based terminology such as ``general purpose'' may be better regulatory terms instead. Furthermore, a leading approach is to to bound ``risky'' AI models in terms of the amount of computation required to train them. As we demonstrate in Section \ref{sec:cape-correlated}, these thresholds do not appropriately bound ``risky'' AI models---a driving goal for regulatory efforts. Additionally, the documents that discuss training on ``large'' amounts of data do not define how many data points meet the bar, leaving leeway for bound parties to argue exemptions.\footnote{Historically, the computing paradigm of ``big data'' suffered from similar criticisms with no concrete amount or volume of data being defined for the purpose of strict regulation.}

\subsection{Capability and Model Size are not Strictly Correlated}
\label{sec:cape-correlated}
Today’s AI governance efforts regularly seek to define frontier models by their size or the amount of computation required to train them. As reflected in some of the governance documents analyzed above, a common strategy is to set a regulatory threshold on the number of parameters included in a model. The rationale behind this approach is a set of experiments that demonstrate that models with larger numbers of parameters, with all other factors held constant, suddenly perform drastically better on downstream tasks they are not explicitly trained for~\citep{weiEmergentAbilitiesLarge2022}. This phenomenon was termed ``emergence'' and drove fears that sufficiently large models, by default, can perform well on tasks that pose risks to public safety.

\begin{mybox}{Parameter}
A variable within a model that is learned from the training data. Parameters define how the model makes predictions by influencing the model's internal structure and decision-making process.
\end{mybox}

Discussions prioritizing model size as a viable threshold fixate on a superficial, easy-to-obtain quantity that is ultimately a red herring. In reality, model capacity and generalizability represent characteristics that are innately difficult to quantify and measure. Not only are current generalization benchmarks lacking in accurate definitions for model capabilities~\citep{rajiAIEverythingWhole2021}, but it is exceedingly common for smaller, more task-focused models to perform better than large, broad-purpose models on specific downstream tasks, as demonstrated below.

\begin{mybox}{Downstream tasks}
Applications that take model outputs as input, re-purposing model knowledge for a new problem that it was not explicitly trained for. The model parameters may optionally be updated via further training using additional data from these tasks. 
\end{mybox}

\begin{figure}[ht]
    \centering
    \begin{minipage}[t]{0.48\textwidth}
        \centering
        \includegraphics[width=\textwidth]{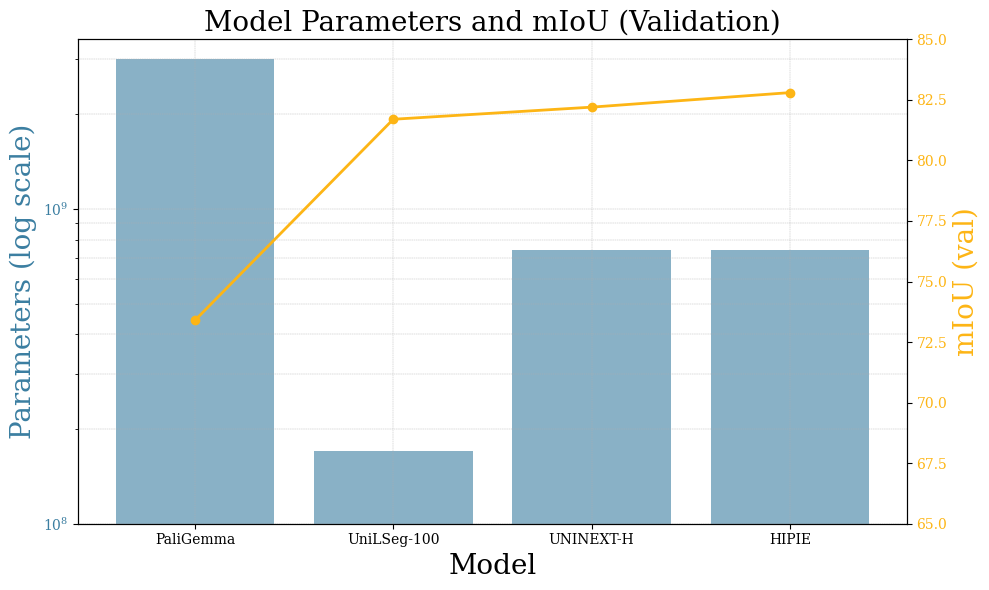}
    \end{minipage}
    \hfill
    \begin{minipage}[t]{0.48\textwidth}
        \centering
        \includegraphics[width=\textwidth]{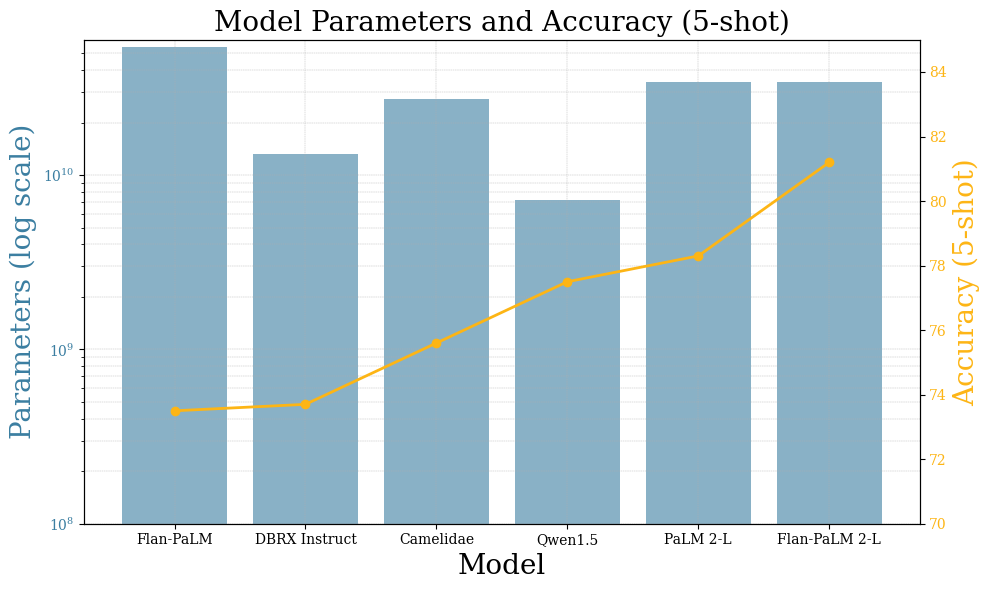}
    \end{minipage}
    \vspace{-0.2cm}
    \caption{
        \textbf{The effectiveness of a model isn't solely determined by its size or computational complexity.} (Left) Despite PaliGemma having an order of magnitude more parameters than UniLSeg, it performs 9.4 mIoU points worse on the common RefCOCO (val) benchmark. (Right) Larger models do not necessarily perform better than smaller ones on the common MMLU benchmark.
    }
    \label{fig:refcoco-mmlu}
\end{figure}

We use the task of image segmentation as an example where smaller models can outperform their larger counterparts. Examples of image segmentation include both civil and national security applications such as building damage assessment or object targeting~\citep{guptaXBDDatasetAssessing2019, guptaOpenSourceAssessmentsAI2024}. Specifically, we examine RefCOCO~\citep{kazemzadehReferItGameReferringObjects2014}, a common image segmentation dataset used to train vision-language models (VLMs), and two models which attain near-state-of-the-art performance on it, PaliGemma~\citep{beyerPaliGemmaVersatile3B2024} and UniLSeg~\citep{liuUniversalSegmentationArbitrary2023}. PaliGemma is a large VLM released openly by Google consisting of $3.0\times10^9$ parameters~\citep{googlePaliGemmaModelCard2024}. On the other hand, UniLSeg, released by Tsinghua University, ByteDance, and the University of Hong Kong, consists of only $1.7\times10^8$ parameters—an order of magnitude smaller than PaliGemma. Yet, UniLSeg achieves a mean intersection-over-union of 81.7 versus PaliGemma’s 73.4 on RefCOCO, which is a massive gain of $\sim$11.3\% in performance.\footnote{Model performance numbers are obtained from their respective papers and \href{https://paperswithcode.com/sota/referring-expression-segmentation-on-refcoco}{Papers With Code}. Parameter counts are derived from the respective papers.} Figure \ref{fig:refcoco-mmlu} additionally demonstrates the performance of two more near-state-of-the-art models, UNINEXT~\citep{yanUniversalInstancePerception2023} and HIPIE~\citep{wangHierarchicalOpenvocabularyUniversal2023}, on RefCOCO for completeness.

\begin{mybox}{Image segmentation}
A task in which an image is split into regions which each represent a specific object type or concept.
\end{mybox}

\begin{mybox}{Mean intersection-over-union}
A measure of how accurately and precisely, on average, an image segmentation model performs. This ranges from zero to one.
\end{mybox}

To further underline this point, we visualize the accuracy of top open-source language models on the Massive Multitask Language Understanding benchmark\footnote{MMLU consists of questions in the ``subjects in the humanities, social sciences, hard sciences, and other areas that are important for some people to learn.''}~\citep{hendrycksMeasuringMassiveMultitask2020} as a function of model parameter count in Figure \ref{fig:refcoco-mmlu}. Low parameter count does not imply incapability, therefore parameter counts alone are an insufficient quantity to define capability frontiers. More parameters are helpful insofar they can fit an appropriately larger amount of data---the two concepts must be bundled to properly circumscribe AI capabilities.

\subsection{A Misplaced Focus on FLOPs}

\begin{mybox}{FLOPs}
Floating point operations. The cumulative number of floating point operations (e.g., 3.5 x 7.1) used during model training. Not to be confused with floating point operations \textit{per second (FLOPS, with a capital S)}.
\end{mybox}

Definitions of foundation and frontier models (see Table \ref{tab:definitions}) include regulatory thresholds defined by cumulative training FLOPs. These numbers have no basis in outcomes or technical reality, as we demonstrate in the following section. $10^{26}$ FLOPs appeared as an arbitrary FLOPs threshold in the October 2023 Biden Administration Executive Order.

Some of the largest models in existence today are sufficient to employ in harmful activities~\citep{openaiDisruptingMaliciousUses2024, openaiDisruptingDeceptiveUses2024}, yet all fail to meet American FLOPs thresholds (see Table \ref{tab:estimated-FLOPs}), raising questions about the threshold's usefulness. These same models are covered under the the EU’s proposed threshold of $10^{25}$ for AI models. However, a fractured environment in which a model regulated in France might not be subject to the same regulations in the United States will lead to confusion.

These thresholds further exacerbate the perception that frontier capabilities can only arise from large models trained with a large amount of computation on larger datasets. As we further demonstrate in this section, even smaller models trained with fewer resources on smaller datasets can set a capability frontier. In fact, research incentives necessitate the creation of methods that reduce computational needs for model training---a trend that is be contrary to regulatory assumptions.

\begin{figure}[ht]
    \centering
    \includegraphics[width=\columnwidth]{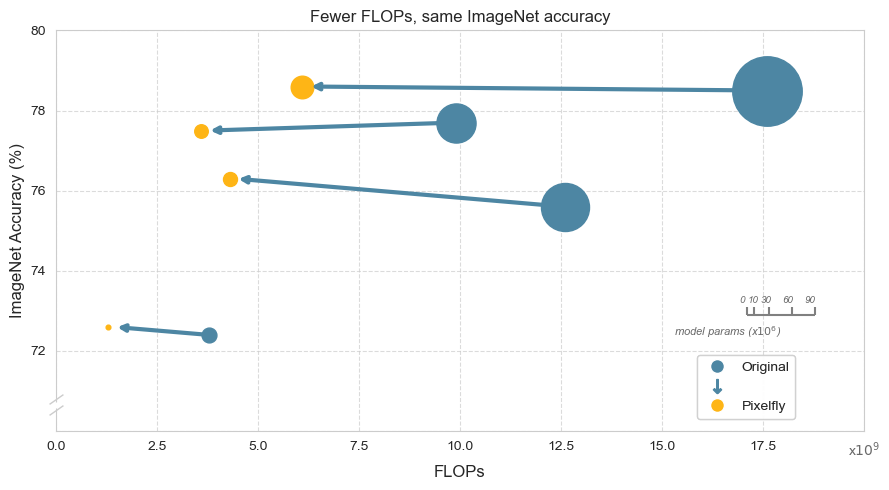}
    \caption{\textbf{Model size and FLOPs are insufficient determinants of capability.} Pixelfly, a recent advancement in efficient model training, can maintain performance on ImageNet across many types of models while reducing their parameter counts and training FLOPs 68\% and 200\% on average, respectively. Each pair of dots represents a Mixer-S/B and ViT-S/B model and its Pixelfly variant.}
    \label{fig:pixelfly}
\end{figure}

\paragraph{Optimizations reverse trends.} One way to visualize the futility of FLOPs thresholds is via recent works such as those on efficient sparse training~\citep{chenPixelatedButterflySimple2021} (Figure \ref{fig:pixelfly}) or other architectural improvements~\citep{zhuScalableMatMulfreeLanguage2024}. They demonstrate that model performance can, in some cases, be decoupled from computational cost---models can train faster and more accurately with fewer parameters and FLOPs. Further research demonstrates decoupling in the opposite direction, i.e., efficient training can occur in compute-constrained environments. Models distributed across multiple machines can be trained with a fraction of parameters while equaling performance at the cost of increased FLOPs~\citep{huhTrainingNeuralNetworks2024}. As such, policies solely relying on FLOP ceilings to bound ``frontier'' models are relying on simplified computing proxies that may not correlate to desired outcomes of controlling the spread of ``risky'' models.

Public disclosure of metrics such as FLOPs is beneficial, however, most well-known commercial AI models do not publicly disclose the amount of FLOPs utilized in the course of training their models. Open-source models, by definition, have exact FLOPs counts available. Below, we provide estimates of FLOPs for a variety of large vision and language models, both commercial and open-source. For proprietary models, these estimates are based on assessments from third-parties rather than concrete disclosures from the respective AI companies.

\begin{table}[ht]
\centering
\renewcommand{\arraystretch}{1.5} % Adjust row height
\arrayrulecolor[HTML]{D3D3D3}
\small
\begin{tabular}{|l|l|l|}
\hline
\rowcolor[HTML]{E0DED3}
Model & Model Type & Estimated FLOPs \\ \hline
Mistral~\citep{jiangMistral7B2023} & Open-source LLM & Training details not disclosed. \\ \hline
LWM~\citep{liuWorldModelMillionLength2024} & Open-source vision model & $5.6\times10^{22}$~\footnote{\label{footnote:direct}Direct communication with paper authors.} \\ \hline
Gemma-7B~\citep{gemmateamGemmaOpenModels2024} & Open-source LLM & $2.5\times10^{23}$~\citep{ruanObservationalScalingLaws2024} \\ \hline
Qwen-72B~\citep{baiQwenTechnicalReport2023} & Open-source LLM & $1.3\times10^{24}$~\citep{rahmanTrackingLargeScaleAI2024} \\ \hline
Falcon-180B~\citep{almazroueiFalconSeriesOpen2023} & Open-source LLM & $3.8\times10^{24}$~\citep{rahmanTrackingLargeScaleAI2024} \\ \hline
Claude-2 & Proprietary LLM & $3.9\times10^{24}$~\citep{rahmanTrackingLargeScaleAI2024} \\ \hline
Llama-3-70B & Open-source LLM & $6.3\times10^{24}$~\citep{rahmanTrackingLargeScaleAI2024} \\ \hline
ChatGPT-4 & Proprietary LLM & $2.2\times10^{25}$~\citep{mcguinnessGPT4DetailsRevealed2023} \\ \hline
Gemini 1.5~\citep{geminiteamGemini15Unlocking2024} & Proprietary LLM & $5.0\times10^{25}$~\citep{rahmanTrackingLargeScaleAI2024} \\ \hline
LVM-3B~\citep{baiSequentialModelingEnables2023} & Open-source vision model & $7.6\times10^{21}$~\footnote{Derivation from OpenLLaMA, which is what LVM is based on, using Chinchilla scaling estimates. Verified with direct communication with the paper authors.} \\ \hline
\end{tabular}
\caption{Large commercial and open-source AI models and their estimated FLOPs.}
\label{tab:estimated-FLOPs}
\end{table}

\paragraph{Efficient methods develop rapidly.} AI research progresses rapidly and the development of efficient methods is an entire subfield with deep financial incentives. The amount of FLOPs needed for a given model architecture to reach a target performance threshold generally tends to drop significantly over a short period of time as the machine learning community identifies software and hardware optimizations for widely-used models. % For example, two recent advancements in efficient model training released in February 2024~\citep{huhTrainingNeuralNetworks2024} and March 2024~\citep{zhaoGaLoreMemoryEfficientLLM2024} were usurped less than three months later (May 2024)~\citep{yangCoMERAComputingMemoryEfficient2024} with drastic reductions in memory usage, wall clock time, and---most relevantly---FLOP count while training. Importantly, these works do not stand alone; they are only the most recent papers following a long-standing, collective effort in efficient model training that advances in lockstep with innovations in model performance and utility.

\begin{mybox}{Transformer}
A widely-used model architecture across vision, language, and other modalities, capable of learning from large datasets and achieving high performance on downstream tasks. 
\end{mybox}

\begin{figure}[ht]
    \centering
    \includegraphics[width=\columnwidth]{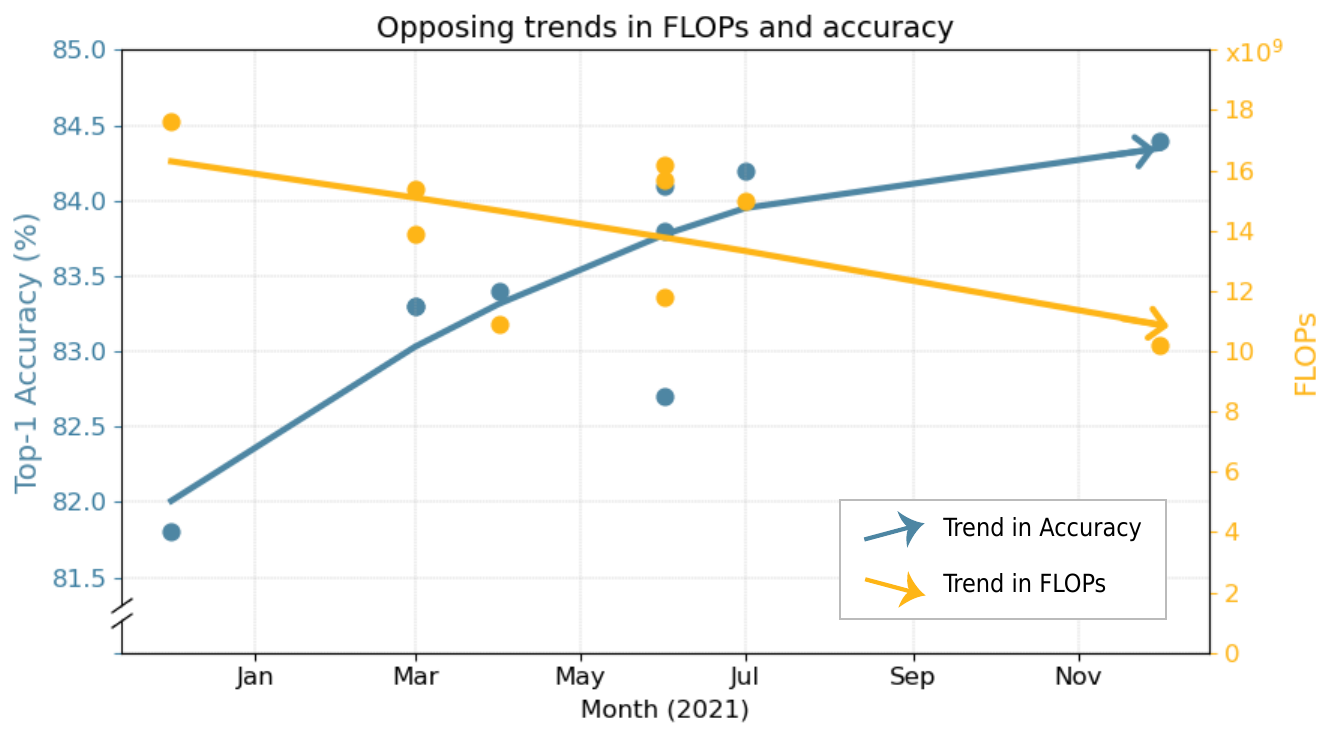}
    \caption{\textbf{Top-1 accuracy and GFLOPs for various models on the ImageNet-1K benchmark.} The rapid pace of development of models results in better performance with fewer FLOPs.}
    \label{fig:FLOPs-in1k}
\end{figure}

To illustrate this point concretely, we consider various vision transformers\footnote{DeiT, PVTv2, CaiT, CoAtNet, XCiT, Swin, MViTv1, MViTv2. Numbers are gathered from the MViTv2 paper and are on models using a comparable amount of computation.} trained on the ImageNet-1K classification benchmark~\citep{russakovskyImageNetLargeScale2014}. In less than a year, the ML research community increased the achieved top-1 accuracy on the benchmark from 81.8\% to 84.4\% while reducing the required FLOPs by 42\% from 17.6 to 10.2 GFLOPs (see Figure \ref{fig:FLOPs-in1k}). This trend holds true for large language models as well~\citep{daoTransformersAreSSMs2024}.

% Efforts to regulate AI based on model size and compute requirements have not kept up with the rapidly advancing pace of AI research which has still not settled on definitions and standards for the field. There exists no clear consensus~\citep{gemaAreWeDone2024, fuBLINKMultimodalLarge2024} amongst ML researchers on comprehensive model evaluation workflows or the metrics that can accurately represent the outputs of such an evaluation. Further complicating governance efforts, performance thresholds such as parameter counts and FLOPs are consistently used as a challenge for aspiring researchers, resulting in a consistent year-over-year reduction in those numbers while achieving equivalent or better capabilities. Focusing on single points of measurement in a rapidly evolving field risks developing brittle policies and an environment where regulators and policymakers are unable to play catch up.

\section{Data is Missing from the Conversation}
\label{sec:data-missing}
Machine learning capabilities are not singularly determined by their model architecture. Rather, machine learning capabilities are defined by \textit{both} the model and the data provided. We define ``data'' as any information a model is exposed to, whether it is during training or deployment. This paper aims to center data in AI governance conversations. We suggest that models alone are not harmful; rather, the unique combination of models exposed to specific datasets (whether during training or inference) \textit{and} subsequently being used for specific purposes may pose a risk to public safety~\citep{baldridgeTerminologyAIRegulation2024}.

Traditionally, training data (both pre-training and fine-tuning) was the only source of information that a model would have access to before making a prediction. However, models can now incorporate new, unseen data during inference through frameworks such as prompting and Retrieval-Augmented Generation (RAG)~\citep{lewisRetrievalAugmentedGenerationKnowledgeIntensive2020}. Therefore, \textit{both} the training and deployment data are relevant when considering how a model incorporates information in its outputs.

\subsection{Big Data to Usable Information}
The rapid rise of AI since approximately 2010 can largely be attributed to (1) advancements in computational hardware in accordance with Moore’s Law, and (2) a focus on large quantities of data. Models are useless without data, and the availability of ``foundational'' datasets, such as ImageNet~\citep{dengImageNetLargescaleHierarchical2009} and Common Crawl,\footnote{\href{https://commoncrawl.org/the-data/ }{https://commoncrawl.org/the-data/}} brought modern machine learning capabilities to bear. Today, AI datasets are often orders of magnitude larger, created by scraping content across the internet.

Dataset size is a key component in ``scaling laws,'' or predictions of performance within a family of models as a function of variables in a training recipe. Research in this area finds strong relationships between model performance and amount of training data, amount of computation, and model parameters~\citep{kaplanScalingLawsNeural2020, hoffmannTrainingComputeOptimalLarge2022, zhaiScalingVisionTransformers2022, googlePaLMTechnicalReport2023}. Additionally, both \citet{hoffmannTrainingComputeOptimalLarge2022} and \citet{googlePaLMTechnicalReport2023} find that model and optimal dataset size scale at \textit{equal proportions} as training compute increases.

However, even an optimal training recipe with an appropriate amount of data, parameters, and compute does not necessarily produce a useful model. The dataset \textit{content} is a crucial factor. A model ``trained on the internet'' can unsurprisingly exhibit the same bias~\citep{fleisigLinguisticBiasChatGPT2024} and toxicity~\citep{liangHolisticEvaluationLanguage2023} present in the data and also fall short in other areas: it may fail at logical reasoning~\citep{berglundReversalCurseLLMs2023}, algebraic computation, or following a user's instructions, to name a few examples. In a limiting argument, a multi-trillion parameter model trained only on Shakespeare novels may never be able to reason about chemical weapon design.

To address this, models are fine-tuned on higher-quality, curated data. Popular techniques that rely on high-quality data include instruction tuning (e.g., reinforcement learning from human feedback, or RLHF~\citep{ouyangTrainingLanguageModels2022}), training models to use tools or act as agents~\citep{schickToolformerLanguageModels2023}, or supervised fine-tuning for a specific task, such as generating images in a particular artistic style.

There is evidence that with the right data and training regime, models in the millions or single digit billions of parameters can perform comparably, if not better, than counterparts orders of magnitude larger in many domains~\citep{yuOpenClosedSmall2023, yuanTinyGPTVEfficientMultimodal2024, eldanTinyStoriesHowSmall2023}. In fact, once a model is sufficiently large, focusing on improving the quality and utilization of data can yield greater gains in performance for a task over simply increasing the size of the model. For example, the Retrieval Augmented Fine-Tuning~\citep{zhangRAFTAdaptingLanguage2024} framework has been shown to improve the question-answering performance of a 7B parameter Llama2 language model over that of GPT-3.5, which otherwise significantly outperforms it out-of-the-box.

\section{Data-Centrism Opens new Analytic Frontiers}
Modern machine learning methods are useful beyond traditional data querying and correlation tools such as search engines in part due to their ability to retrieve, compile, and organize data even when given unspecific queries. Below we outline two distinct features that are uniquely enabled by the combination of ML models and data: (1) \textit{retrieval}, where a model outputs information retrieved directly from its data, and (2) \textit{derivation}, where a model compiles or synthesizes items from provided data to generate new information. These features enable new ways to interact with complex data that would otherwise be difficult to manage, offering potential benefits as well as risks, which we explore further below.

\subsection{Retrieval}
As our ability to collect and maintain digital information has soared over the last few decades, retrieving the right results for a certain query has become a core technological focus. Billions of dollars have been spent towards developing efficient data representations for search engines~\citep{brinAnatomyLargescaleHypertextual1998, deanMapReduceSimplifiedData2008} and databases~\citep{corbettSpannerGooglesGloballyDistributed2012, shvachkoHadoopDistributedFile2010}, and towards creating the algorithms to find and retrieve these results. Now, AI models trained on large amounts of data have become both capable encoders and retrievers of data (in addition to generators, as we describe in the section on derivation). This becomes a problem when a model has been exposed to specific data points that would be considered sensitive if directly retrieved, such as credit card numbers or classified information. The retrieval itself could occur either through (1) a model memorizing and then reproducing points in training data, or (2) retrieving from a large amount of data provided at test time, such as a company's internal database. Below, we describe these two cases in more detail.

\begin{mybox}{Retrieval}
A task requiring models to accurately locate and return a requested piece of information. 
\end{mybox}

\paragraph{Retrieval from training data.} Datasets contaminated with outliers have historically relied on dataset volume to dilute outlier effects. This leads to the misconception that a small quantity of ``harmful'' data points can be negated by massive amounts of otherwise commonplace data. Unfortunately, this intuition does not translate to modern machine learning methods. Large models are known to memorize some parts of training data and reproduce them if queried correctly~\citep{carliniQuantifyingMemorizationNeural2022}. Therefore, large models can retrieve, and therefore utilize, harmful data even if it is present in a negligible quantity.

However, memorization does not occur across data equally: prior work shows that ``average'' training samples are less likely to be memorized, whereas outlierse and duplicated data points are more likely to be memorized~\citep{feldmanWhatNeuralNetworks2020, feldmanDoesLearningRequire2020, carliniQuantifyingMemorizationNeural2022}. As certain types of data, such as child sexual abuse material, are outliers on the Internet~\citep{thielIdentifyingEliminatingCSAM2023}, memorization of such data poses an inherent risk in the downstream usage of affected models, especially combined with the powerful retrieval abilities of current models. \citet{nasrScalableExtractionTraining2023} is another example of work where ChatGPT was used to retrieve training data which comprised personally identifiable information of dozens of individuals.

\paragraph{Retrieval from previously unseen data.} An AI system may also be exposed to entire new domains of data during inference that were not present during training. Models can utilize new, unseen data through prompting or integration with external databases. Models' ability to effectively interpret new domains without prior training marks a significant shift in how we store and use information. Instead of creating expensive, specialized systems to process data like financial documents or hospital records, modern general-purpose models can understand and work with novel data formats they have never encountered before while requiring minimal engineering effort.

Many of today’s large models are being specifically designed to respond flexibly to new tasks and prompt formats. Concretely, in-context learning~\citep{brownLanguageModelsAre2020} (ICL) allows users to provide example input-output pairs of a task to a large model which can equip it to solve novel instances of that task. ICL can, for example, pose a risk to society in the automation of attack prompt generation, achieved by instructing LLMs to mimic high-quality human-crafted prompts in large volumes~\citep{dengAttackPromptGeneration2023}. Further, modern AI systems may be used to efficiently sift through large amounts of data at inference time---even if they have not seen it before---using frameworks such as Retrieval Augmented Generation (RAG)~\citep{gaoRetrievalAugmentedGenerationLarge2024}. In RAG, given a user query, an answer can be generated by efficiently searching a database for relevant concepts and making sense of this new information to return a relevant response~\citep{yasunagaRetrievalAugmentedMultimodalLanguage2023, kongAudioFlamingoNovel2024, blattmannRetrievalAugmentedDiffusionModels2022}. As these systems can now return sensitive examples not seen before by dynamically augmenting their knowledge or understanding new tasks from test-time examples, they can therefore be used in the furtherance of actions that pose risks to society by drastically lowering the boundary to both finding and exploiting risk-posing information~\citep{barrettIdentifyingMitigatingSecurity2023}.

\subsection{Derivation}
As AI capabilities increase, a growing concern is the generation of original or derivative information that is more revealing that the data provided to the model initially. For example, if a system is given two entry-level textbooks in physics and chemistry, respectively, and uses independent concepts from either to build a toy rocket, we would call the process of arriving at the toy rocket instructions ``derivation.''

This feature is especially present in modern machine learning methods when compared to technologies such as databases due to their ability to synthesize unrelated pieces of information on the fly. While retrieved content is often straightforward to recognize and check---i.e., it may be quickly obvious that a generated phone number is real, and possible to check if a particular image was contained within training or deployment-time data---derived content is more nuanced and difficult to measure, and thus may present a greater concern.

\begin{mybox}{Derivation}
When a model compiles disparate pieces of data to infer novel piece of information not explicitly contained in any of its components.
\end{mybox}

Under this category, multiple pieces of otherwise mundane information could be compiled to form information that is now sensitive. For instance, a language model trained for code generation could be provided a description of a vulnerability and be used to generate code for exploiting it. Models have already begun to present synthesis capabilities in different arenas, such as for code generation of programming languages with low data availability~\citep{moraSyntheticProgrammingElicitation2024} and the generation of Mathematics Olympiad-level geometric proofs as part of larger pipelines~\citep{trinhSolvingOlympiadGeometry2024}.

The maximal extent to which current models are capable of derivation is not yet clear as methodologies for inducing such capabilities are constantly evolving. For example, although modern language models have shown nascent indicators of capability to generate novel research ideas in fields such as natural language processing, the ideas they generate lack diversity and may not be tractable~\citep{siCanLLMsGenerate2024}. Modern image generation models struggle to synthesize images precisely adhering to descriptions of unique combinations of objects and their attributes previously unseen in training data~\citep{huangT2ICompBenchComprehensiveBenchmark2023}. Our intent in this section is not to establish a measure for models’ derivation capability but rather to bring attention to derivation as a unique capability offered by modern ML models.

\section{Assumptions and Limitations}
Given the rapid pace of AI development, we acknowledge the limits of our core analytic assumptions, grounded in the current state-of-art in the field, that drive the analysis and recommendations in this work. If these building blocks are outpaced by future developments, then this work should be revisited.

\paragraph{Assumption 1: Powerful models are unable to reason without memorizing information.} 
Large models can perform well by both learning generalizable semantics over their training data, but also through the rote memorization of data or concepts. Currently, there exists no class of powerful machine learning models which are able to ``reason'' about the world without having memorized any data during its training period. Put another way, there are no reasoning agents that are derived in a manner that is completely detached from data. One can argue that such a model, should it exist, would fit the definition of ``artificial general intelligence'' as it could generalize to any new set of data without inherent data priors.

\paragraph{Assumption 2: Dataset distillation methods are still over the horizon.}
The field’s understanding of the amount of data points needed for a model to achieve proficiency on specific tasks is still evolving. This area of research is termed ``dataset distillation'' and aims to reduce the number of data points necessary to achieve target metrics~\citep{wangDatasetDistillation2020, zhaoDatasetCondensationGradient2021}. Further, it remains unclear what exactly constitutes a ``data point,'' especially with modern methods like transformers, which rely on tokens, the amount of which varies with different tokenization methods~\citep{sennrichNeuralMachineTranslation2016, schusterJapaneseKoreanVoice2012}. We aim to establish one rigorous definition of ``data point'' in future work, as well as analysis of how many data points define emergent capability.

In a limiting argument, should data distillation methods improve to the point where models can learn generalizable knowledge without any data at all, this work would need to be revisited.

\section{Avenues for Data-Forward Regulation}
Given our analysis above, the inclusion of data in nascent AI governance conversations can simplify the regulatory overhead by enabling the use of existing legal frameworks and the creation and execution of novel, data-backed evaluation schemes. Specifically, there are numerous policies and laws surrounding the appropriate use of data in contexts that are deemed to be of risk to the public. Instead of reinventing these policies using a new set of definitions that are model-specific, expanding and modifying them to account for the use of data by powerful models might offer a simpler path towards effective evaluation frameworks in areas where definitions alone are vague, leading to simpler regulations.

\subsection{Applying Existing Data-Focused Legal and Regulatory Approaches}
Significant work has been and continues to be done to mitigate malicious model outputs or behaviors. Thus far, model creators have relied on identifying malicious outputs or behaviors through red teaming and safety training~\citep{ganguliRedTeamingLanguage2022, weiJailbrokenHowDoes2023}.

However, some classes of outputs or behaviors that are deemed risky could more easily be stemmed by careful curation of datasets. Unique information such as the relationship between a person and their social security number, or specific instances of child sexual abuse material, is extremely unlikely to be generated if that data is never provided to a model.

There exists a range of legal and regulatory frameworks that cover many categories of model outputs that are of greatest concern, including personal identifiable information, child sexual abuse material, and classified content. Data-centrism prevents models from acquiring the capacity for harmful behaviors prior to the expenditure of computation. Since existing regulations can be applied, AI governance can be achieved without the need for new regulatory frameworks.

\subsection{Technological Levers for Data-Forward Regulation}
Although research has shown that certain model capabilities emerge once sufficient model size and compute are attained~\citep{weiEmergentAbilitiesLarge2022}, establishing regulatory thresholds is ill-defined given just these two metrics. As discussed in Section \ref{sec:data-missing}, models provided with the right data can perform comparably to, if not better than, larger and more compute intensive alternatives. Further, a model must first be paired with sensitive information for it to make use of it. That is, the model does not exist in a vacuum, and a data-forward approach that prioritizes data content and quality filtration over model size and computation could yield greater benefits in mitigating risks posed by the use of models. Here, we briefly outline examples of existing techniques and argue for the development of new methods.

\paragraph{Existing data filtration.}
Modern web-scale datasets are extremely large, numbering in the billions to trillions of data points. As such, human review of every data point is not possible from either a labor or monetary perspective. However, the volume of data does not permit the abdication of responsibility or duty to curate datasets responsibly. In response, methods have been proposed to partially or fully automate the filtration process~\citep{albalakSurveyDataSelection2024}. Content can be filtered based on fixed patterns such as blacklisted source URLs or key words, however, these methods can be rigid and insensitive to the nuance of usage context. Large vision and language models such as CLIP~\citep{schramowskiCanMachinesHelp2022} and Meta’s Llama Guard~\citep{inanLlamaGuardLLMbased2023} have been used to classify whether data points are risky under human-defined criteria and can be more sensitive to context than blacklist-based methods. However, these methods are far from perfect---offering an important avenue for future research.

\paragraph{Quantifying risk for workloads.}
In addition to data filtration schemes, a rigorous evaluation framework for powerful AI models that is inclusive of both models and data is needed. Many approaches are feasible, and we detail an evaluation framework under development that attempts to solidify this discussion into a quantifiable benchmark.

For example, imagine asking a model a question in a setting where accuracy of the answer matters, say ``what materials make up Saturn’s rings?'' Short, broken answers such as ``rock, water'' would be regarded as unreliable or incorrect as opposed to an answer that demonstrates mastery of grammar and facts such as ``The rings of Saturn are primarily composed of countless small particles of ice and rock. These particles range in size from tiny grains of dust to larger chunks that can be several meters across.''

For a specific type of output, there is likely a minimum size threshold for a model to be capable of learning the syntax of that output domain~\citep{chenSuddenDropsLoss2024}. The initial stage of model training is focused on acquiring \textit{fluency}---object detection models learn what the shape and proportion of a valid detection looks like, and language models learn the underlying structure and grammar of the languages over which they operate. In this stage, models are parameter-bound---the largest gains in fluency are likely to come from making models bigger. However, once a model has passed this hypothesized stage to learn the syntax ``well enough,'' we posit that the model is now data-bound and improvements to performance, or correctness, are more likely to come from improvements to the content and utilization of data rather than just from arbitrary scaling~\citep{weiEmergentAbilitiesLarge2022, yuOpenClosedSmall2023, eldanTinyStoriesHowSmall2023}.

This inherent relationship between fluency and correctness can be used as a powerful tool to regulate AI capabilities in a data-parameter inclusive fashion. For any arbitrary task, the performance of a model on that task can be plotted on a fluency-correctness curve. Once all workloads are plotted, the resulting risk profile can be adjudicated and a resulting judgment---whether a reduction in the parameter count of the model or a specific pruning of the training dataset is necessary---can be made by the model developers.

\begin{mybox}{Fluency and correctness are somewhat dependent}
Fluency and correctness cannot be plotted as two independent axes. Since semantics are somewhat dependent on syntax for many languages, these concepts can sometimes be used interchangeably.
\end{mybox}

Ultimately, such an evaluation framework can aid in the development of regulatory system through which the government and model developers can safely, privately, and precisely iterate on removing the ability of models to aid in risky tasks prior to model release.

\subsection{Incentivizing Data Governance Tools and Practices}
Just as existing policies and regulations advocate for the standardization of model documentation, such as model or system cards~\citep{mitchellModelCardsModel2019}, data-centrism motivates the standardization of dataset documentation. Comprehensive approaches for doing so have been proposed already, such as Datasheets for Datasets~\citep{gebruDatasheetsDatasets2021} or Data Cards~\citep{pushkarnaDataCardsPurposeful2022}. These documentation formalisms currently detail dataset properties regarding content, structure, preprocessing, distribution, and intended or potential use cases. Given the common practice of aggregating datasets from multiple sources, mechanisms for documenting and tracking the provenance of dataset contents, such as Data Provenance Cards~\citep{longpreDataProvenanceInitiative2023}, would greatly ease verification of information available to a model. Further, standardized ontologies~\citep{zengAIRiskCategorization2024} that categorize and rank ``risky information'' can be applied to each dataset in a provenance card, which can then be used as a first approximation of potential retrieval and derivation capabilities.

Red teaming~\citep{perezRedTeamingLanguage2022, baiConstitutionalAIHarmlessness2022} has hitherto emerged as the \textit{de facto} standard to evaluate the potential of models intended for release to pose a risk to public safety. However, red teaming is, as of yet, not standardized. Further, with the rapid increase in the amount of models that need to be assessed, there exists no mechanism through which the potential of models to perform specific tasks can be estimated \textit{a priori}. The development of a technical framework for measuring the dynamics of the performance of a model family for a given task as a function of both model scale and quality of training data, particularly one that can identify inflection points at which a model’s performance becomes \textit{bound by its data} rather than its size, would be an important tool for more precisely identifying when models could feasibly be used in the furtherance of behaviors that harm society.

\section{Conclusion: Evolving AI Governance Alongside AI Technology}
Despite rapid growths in both model and dataset sizes in recent years, AI policies have hinged on thresholds, definitional concepts, and qualifiers that limit their medium-to-long term liability. For a technology that will be with us for the foreseeable future, we can, and should, approach governance in a more deliberate manner, with a clear understanding of what enables these capabilities to be powerful in the first place.

Similar to how an arbitrarily large engine, no matter how specifically quantified, would be useless without defining the kind of fuel used with it, the AI policy landscape mistakenly focuses on a small set of model-based thresholds, particularly FLOP and parameter counts. Neither fully define how powerful a machine learning model may be without an understanding of the data that accompanies them. Furthermore, the lack of definitional clarity with what constitutes a ``frontier'', ``foundation'', ``dual-use'', or ``general purpose'' model complicates governance efforts. More generally, these two trends in governance further propagate the outdated idea that the largest, most compute intensive models are those which drive AI risk. As we reach a point where smaller models, when paired with large, foundational datasets or small, high-quality datasets, can perform as well as larger models, this narrow approach creates loopholes and unfairly penalizes otherwise beneficial technologies.

Centering data offers a more durable approach to AI governance, particularly as trends in quantifiable measures of model capability are difficult to predict. A focus on data also provides an opportunity to better research, define, and respond to benefits and risks posed by AI, a debate that remains nebulous in both policy and technical circles. Centering data also provides avenues for existing regulations surrounding sensitive types of data to apply while also clearing the way for new evaluation methods to quantify the use of data and models together. Expanding model-based regulations to focus additionally on their paired data builds a stronger foundation that is less prone to collapse.

While a pivot in the governance landscape may be daunting, a focus on data provides the opportunities and incentives for government, academic researchers, civil society, and the private sector to develop new tools and approaches that lead to meaningful policies. This paper is the first of the Frontier Data Initiative which seeks to focus governance efforts on the combination of data and models. Future papers in this series will focus on novel technical approaches to benchmarking dataset capabilities, auditing data and conducting data forward red teaming. These technical papers will be joined by policy papers centering data-forward AI governance opportunities.

\section*{Acknowledgements}
We are thankful to many friends, colleagues, and collaborators for the numerous discussions, critiques, and edits to this paper including Yutong Bai, David Chan, Lisa Dunlap, Michelle Li, Sanjeev Raja, Courtney Rankin, Anand Siththaranjan, and Sanjay Subramanian. Authors, as part of their affiliation with UC Berkeley, were supported in part by the National Science Foundation, U.S. Department of Defense, Founders Pledge, Ford Foundation, and/or the Berkeley Artificial Intelligence Research (BAIR) industrial alliance program.

\clearpage

\bibliography{iclr2025_conference}

\begin{thebibliography}{101}
\providecommand{\natexlab}[1]{#1}
\providecommand{\url}[1]{\texttt{#1}}
\expandafter\ifx\csname urlstyle\endcsname\relax
  \providecommand{\doi}[1]{doi: #1}\else
  \providecommand{\doi}{doi: \begingroup \urlstyle{rm}\Url}\fi

\bibitem[Albalak et~al.(2024)Albalak, Elazar, Xie, Longpre, Lambert, Wang, Muennighoff, Hou, Pan, Jeong, Raffel, Chang, Hashimoto, and Wang]{albalakSurveyDataSelection2024}
Alon Albalak, Yanai Elazar, Sang~Michael Xie, Shayne Longpre, Nathan Lambert, Xinyi Wang, Niklas Muennighoff, Bairu Hou, Liangming Pan, Haewon Jeong, Colin Raffel, Shiyu Chang, Tatsunori Hashimoto, and William~Yang Wang.
\newblock A {{Survey}} on {{Data Selection}} for {{Language Models}}, February 2024.

\bibitem[Almazrouei et~al.(2023)Almazrouei, Alobeidli, Alshamsi, Cappelli, Cojocaru, Debbah, Goffinet, Hesslow, Launay, Malartic, Mazzotta, Noune, Pannier, and Penedo]{almazroueiFalconSeriesOpen2023}
Ebtesam Almazrouei, Hamza Alobeidli, Abdulaziz Alshamsi, Alessandro Cappelli, Ruxandra Cojocaru, M{\'e}rouane Debbah, {\'E}tienne Goffinet, Daniel Hesslow, Julien Launay, Quentin Malartic, Daniele Mazzotta, Badreddine Noune, Baptiste Pannier, and Guilherme Penedo.
\newblock The {{Falcon Series}} of {{Open Language Models}}.
\newblock https://arxiv.org/abs/2311.16867v2, November 2023.

\bibitem[Alstott(2023)]{alstottPreparingFederalResponse2023}
Jeff Alstott.
\newblock Preparing the {{Federal Response}} to {{Advanced Technologies}}.
\newblock Technical report, RAND Corporation, September 2023.

\bibitem[Anderljung et~al.(2023)Anderljung, Barnhart, Korinek, Leung, O'Keefe, Whittlestone, Avin, Brundage, Bullock, {Cass-Beggs}, Chang, Collins, Fist, Hadfield, Hayes, Ho, Hooker, Horvitz, Kolt, Schuett, Shavit, Siddarth, Trager, and Wolf]{anderljungFrontierAIRegulation2023}
Markus Anderljung, Joslyn Barnhart, Anton Korinek, Jade Leung, Cullen O'Keefe, Jess Whittlestone, Shahar Avin, Miles Brundage, Justin Bullock, Duncan {Cass-Beggs}, Ben Chang, Tantum Collins, Tim Fist, Gillian Hadfield, Alan Hayes, Lewis Ho, Sara Hooker, Eric Horvitz, Noam Kolt, Jonas Schuett, Yonadav Shavit, Divya Siddarth, Robert Trager, and Kevin Wolf.
\newblock Frontier {{AI Regulation}}: {{Managing Emerging Risks}} to {{Public Safety}}, November 2023.

\bibitem[Bai et~al.(2023{\natexlab{a}})Bai, Bai, Chu, Cui, Dang, Deng, Fan, Ge, Han, Huang, Hui, Ji, Li, Lin, Lin, Liu, Liu, Lu, Lu, Ma, Men, Ren, Ren, Tan, Tan, Tu, Wang, Wang, Wang, Wu, Xu, Xu, Yang, Yang, Yang, Yang, Yao, Yu, Yuan, Yuan, Zhang, Zhang, Zhang, Zhang, Zhou, Zhou, Zhou, and Zhu]{baiQwenTechnicalReport2023}
Jinze Bai, Shuai Bai, Yunfei Chu, Zeyu Cui, Kai Dang, Xiaodong Deng, Yang Fan, Wenbin Ge, Yu~Han, Fei Huang, Binyuan Hui, Luo Ji, Mei Li, Junyang Lin, Runji Lin, Dayiheng Liu, Gao Liu, Chengqiang Lu, Keming Lu, Jianxin Ma, Rui Men, Xingzhang Ren, Xuancheng Ren, Chuanqi Tan, Sinan Tan, Jianhong Tu, Peng Wang, Shijie Wang, Wei Wang, Shengguang Wu, Benfeng Xu, Jin Xu, An~Yang, Hao Yang, Jian Yang, Shusheng Yang, Yang Yao, Bowen Yu, Hongyi Yuan, Zheng Yuan, Jianwei Zhang, Xingxuan Zhang, Yichang Zhang, Zhenru Zhang, Chang Zhou, Jingren Zhou, Xiaohuan Zhou, and Tianhang Zhu.
\newblock Qwen {{Technical Report}}, September 2023{\natexlab{a}}.

\bibitem[Bai et~al.(2022)Bai, Kadavath, Kundu, Askell, Kernion, Jones, Chen, Goldie, Mirhoseini, McKinnon, Chen, Olsson, Olah, Hernandez, Drain, Ganguli, Li, {Tran-Johnson}, Perez, Kerr, Mueller, Ladish, Landau, Ndousse, Lukosuite, Lovitt, Sellitto, Elhage, Schiefer, Mercado, DasSarma, Lasenby, Larson, Ringer, Johnston, Kravec, Showk, Fort, Lanham, {Telleen-Lawton}, Conerly, Henighan, Hume, Bowman, {Hatfield-Dodds}, Mann, Amodei, Joseph, McCandlish, Brown, and Kaplan]{baiConstitutionalAIHarmlessness2022}
Yuntao Bai, Saurav Kadavath, Sandipan Kundu, Amanda Askell, Jackson Kernion, Andy Jones, Anna Chen, Anna Goldie, Azalia Mirhoseini, Cameron McKinnon, Carol Chen, Catherine Olsson, Christopher Olah, Danny Hernandez, Dawn Drain, Deep Ganguli, Dustin Li, Eli {Tran-Johnson}, Ethan Perez, Jamie Kerr, Jared Mueller, Jeffrey Ladish, Joshua Landau, Kamal Ndousse, Kamile Lukosuite, Liane Lovitt, Michael Sellitto, Nelson Elhage, Nicholas Schiefer, Noemi Mercado, Nova DasSarma, Robert Lasenby, Robin Larson, Sam Ringer, Scott Johnston, Shauna Kravec, Sheer~El Showk, Stanislav Fort, Tamera Lanham, Timothy {Telleen-Lawton}, Tom Conerly, Tom Henighan, Tristan Hume, Samuel~R. Bowman, Zac {Hatfield-Dodds}, Ben Mann, Dario Amodei, Nicholas Joseph, Sam McCandlish, Tom Brown, and Jared Kaplan.
\newblock Constitutional {{AI}}: {{Harmlessness}} from {{AI Feedback}}, December 2022.

\bibitem[Bai et~al.(2023{\natexlab{b}})Bai, Geng, Mangalam, Bar, Yuille, Darrell, Malik, and Efros]{baiSequentialModelingEnables2023}
Yutong Bai, Xinyang Geng, Karttikeya Mangalam, Amir Bar, Alan Yuille, Trevor Darrell, Jitendra Malik, and Alexei~A. Efros.
\newblock Sequential {{Modeling Enables Scalable Learning}} for {{Large Vision Models}}, December 2023{\natexlab{b}}.

\bibitem[Baldridge et~al.(2024)Baldridge, Coleman, and Sandhu]{baldridgeTerminologyAIRegulation2024}
David Baldridge, Beth Coleman, and Jamie~Amarat Sandhu.
\newblock The terminology of {{AI}} regulation: {{Preventing}} ``harm'' and mitigating ``risk''.
\newblock https://srinstitute.utoronto.ca/news/terminology-regulation-risk-harm, February 2024.

\bibitem[Barrett et~al.(2023)Barrett, Boyd, Burzstein, Carlini, Chen, Choi, Chowdhury, Christodorescu, Datta, Feizi, Fisher, Hashimoto, Hendrycks, Jha, Kang, Kerschbaum, Mitchell, Mitchell, Ramzan, Shams, Song, Taly, and Yang]{barrettIdentifyingMitigatingSecurity2023}
Clark Barrett, Brad Boyd, Elie Burzstein, Nicholas Carlini, Brad Chen, Jihye Choi, Amrita~Roy Chowdhury, Mihai Christodorescu, Anupam Datta, Soheil Feizi, Kathleen Fisher, Tatsunori Hashimoto, Dan Hendrycks, Somesh Jha, Daniel Kang, Florian Kerschbaum, Eric Mitchell, John Mitchell, Zulfikar Ramzan, Khawaja Shams, Dawn Song, Ankur Taly, and Diyi Yang.
\newblock Identifying and {{Mitigating}} the {{Security Risks}} of {{Generative AI}}.
\newblock \emph{Foundations and Trends{\textregistered} in Privacy and Security}, 6\penalty0 (1):\penalty0 1--52, 2023.
\newblock ISSN 2474-1558, 2474-1566.
\newblock \doi{10.1561/3300000041}.

\bibitem[Berglund et~al.(2023)Berglund, Tong, Kaufmann, Balesni, Stickland, Korbak, and Evans]{berglundReversalCurseLLMs2023}
Lukas Berglund, Meg Tong, Maximilian Kaufmann, Mikita Balesni, Asa~Cooper Stickland, Tomasz Korbak, and Owain Evans.
\newblock The {{Reversal Curse}}: {{LLMs}} trained on ``{{A}} is {{B}}'' fail to learn ``{{B}} is {{A}}''.
\newblock In \emph{The {{Twelfth International Conference}} on {{Learning Representations}}}, October 2023.

\bibitem[Beyer et~al.(2024)Beyer, Steiner, Pinto, Kolesnikov, Wang, Salz, Neumann, Alabdulmohsin, Tschannen, Bugliarello, Unterthiner, Keysers, Koppula, Liu, Grycner, Gritsenko, Houlsby, Kumar, Rong, Eisenschlos, Kabra, Bauer, Bo{\v s}njak, Chen, Minderer, Voigtlaender, Bica, Balazevic, Puigcerver, Papalampidi, Henaff, Xiong, Soricut, Harmsen, and Zhai]{beyerPaliGemmaVersatile3B2024}
Lucas Beyer, Andreas Steiner, Andr{\'e}~Susano Pinto, Alexander Kolesnikov, Xiao Wang, Daniel Salz, Maxim Neumann, Ibrahim Alabdulmohsin, Michael Tschannen, Emanuele Bugliarello, Thomas Unterthiner, Daniel Keysers, Skanda Koppula, Fangyu Liu, Adam Grycner, Alexey Gritsenko, Neil Houlsby, Manoj Kumar, Keran Rong, Julian Eisenschlos, Rishabh Kabra, Matthias Bauer, Matko Bo{\v s}njak, Xi~Chen, Matthias Minderer, Paul Voigtlaender, Ioana Bica, Ivana Balazevic, Joan Puigcerver, Pinelopi Papalampidi, Olivier Henaff, Xi~Xiong, Radu Soricut, Jeremiah Harmsen, and Xiaohua Zhai.
\newblock {{PaliGemma}}: {{A}} versatile {{3B VLM}} for transfer, July 2024.

\bibitem[Blattmann et~al.(2022)Blattmann, Rombach, Oktay, M{\"u}ller, and Ommer]{blattmannRetrievalAugmentedDiffusionModels2022}
Andreas Blattmann, Robin Rombach, Kaan Oktay, Jonas M{\"u}ller, and Bj{\"o}rn Ommer.
\newblock Retrieval-{{Augmented Diffusion Models}}.
\newblock \emph{Advances in Neural Information Processing Systems}, 35:\penalty0 15309--15324, December 2022.

\bibitem[Bommasani et~al.(2022)Bommasani, Hudson, Adeli, Altman, Arora, {von Arx}, Bernstein, Bohg, Bosselut, Brunskill, Brynjolfsson, Buch, Card, Castellon, Chatterji, Chen, Creel, Davis, Demszky, Donahue, Doumbouya, Durmus, Ermon, Etchemendy, Ethayarajh, {Fei-Fei}, Finn, Gale, Gillespie, Goel, Goodman, Grossman, Guha, Hashimoto, Henderson, Hewitt, Ho, Hong, Hsu, Huang, Icard, Jain, Jurafsky, Kalluri, Karamcheti, Keeling, Khani, Khattab, Koh, Krass, Krishna, Kuditipudi, Kumar, Ladhak, Lee, Lee, Leskovec, Levent, Li, Li, Ma, Malik, Manning, Mirchandani, Mitchell, Munyikwa, Nair, Narayan, Narayanan, Newman, Nie, Niebles, Nilforoshan, Nyarko, Ogut, Orr, Papadimitriou, Park, Piech, Portelance, Potts, Raghunathan, Reich, Ren, Rong, Roohani, Ruiz, Ryan, R{\'e}, Sadigh, Sagawa, Santhanam, Shih, Srinivasan, Tamkin, Taori, Thomas, Tram{\`e}r, Wang, Wang, Wu, Wu, Wu, Xie, Yasunaga, You, Zaharia, Zhang, Zhang, Zhang, Zhang, Zheng, Zhou, and Liang]{bommasaniOpportunitiesRisksFoundation2022}
Rishi Bommasani, Drew~A. Hudson, Ehsan Adeli, Russ Altman, Simran Arora, Sydney {von Arx}, Michael~S. Bernstein, Jeannette Bohg, Antoine Bosselut, Emma Brunskill, Erik Brynjolfsson, Shyamal Buch, Dallas Card, Rodrigo Castellon, Niladri Chatterji, Annie Chen, Kathleen Creel, Jared~Quincy Davis, Dora Demszky, Chris Donahue, Moussa Doumbouya, Esin Durmus, Stefano Ermon, John Etchemendy, Kawin Ethayarajh, Li~{Fei-Fei}, Chelsea Finn, Trevor Gale, Lauren Gillespie, Karan Goel, Noah Goodman, Shelby Grossman, Neel Guha, Tatsunori Hashimoto, Peter Henderson, John Hewitt, Daniel~E. Ho, Jenny Hong, Kyle Hsu, Jing Huang, Thomas Icard, Saahil Jain, Dan Jurafsky, Pratyusha Kalluri, Siddharth Karamcheti, Geoff Keeling, Fereshte Khani, Omar Khattab, Pang~Wei Koh, Mark Krass, Ranjay Krishna, Rohith Kuditipudi, Ananya Kumar, Faisal Ladhak, Mina Lee, Tony Lee, Jure Leskovec, Isabelle Levent, Xiang~Lisa Li, Xuechen Li, Tengyu Ma, Ali Malik, Christopher~D. Manning, Suvir Mirchandani, Eric Mitchell, Zanele Munyikwa, Suraj Nair,
  Avanika Narayan, Deepak Narayanan, Ben Newman, Allen Nie, Juan~Carlos Niebles, Hamed Nilforoshan, Julian Nyarko, Giray Ogut, Laurel Orr, Isabel Papadimitriou, Joon~Sung Park, Chris Piech, Eva Portelance, Christopher Potts, Aditi Raghunathan, Rob Reich, Hongyu Ren, Frieda Rong, Yusuf Roohani, Camilo Ruiz, Jack Ryan, Christopher R{\'e}, Dorsa Sadigh, Shiori Sagawa, Keshav Santhanam, Andy Shih, Krishnan Srinivasan, Alex Tamkin, Rohan Taori, Armin~W. Thomas, Florian Tram{\`e}r, Rose~E. Wang, William Wang, Bohan Wu, Jiajun Wu, Yuhuai Wu, Sang~Michael Xie, Michihiro Yasunaga, Jiaxuan You, Matei Zaharia, Michael Zhang, Tianyi Zhang, Xikun Zhang, Yuhui Zhang, Lucia Zheng, Kaitlyn Zhou, and Percy Liang.
\newblock On the {{Opportunities}} and {{Risks}} of {{Foundation Models}}, July 2022.

\bibitem[Brin \& Page(1998)Brin and Page]{brinAnatomyLargescaleHypertextual1998}
Sergey Brin and Lawrence Page.
\newblock The anatomy of a large-scale hypertextual {{Web}} search engine.
\newblock \emph{Computer Networks and ISDN Systems}, 30\penalty0 (1):\penalty0 107--117, April 1998.
\newblock ISSN 0169-7552.
\newblock \doi{10.1016/S0169-7552(98)00110-X}.

\bibitem[Brown et~al.(2020)Brown, Mann, Ryder, Subbiah, Kaplan, Dhariwal, Neelakantan, Shyam, Sastry, Askell, Agarwal, {Herbert-Voss}, Krueger, Henighan, Child, Ramesh, Ziegler, Wu, Winter, Hesse, Chen, Sigler, Litwin, Gray, Chess, Clark, Berner, McCandlish, Radford, Sutskever, and Amodei]{brownLanguageModelsAre2020}
Tom Brown, Benjamin Mann, Nick Ryder, Melanie Subbiah, Jared~D Kaplan, Prafulla Dhariwal, Arvind Neelakantan, Pranav Shyam, Girish Sastry, Amanda Askell, Sandhini Agarwal, Ariel {Herbert-Voss}, Gretchen Krueger, Tom Henighan, Rewon Child, Aditya Ramesh, Daniel Ziegler, Jeffrey Wu, Clemens Winter, Chris Hesse, Mark Chen, Eric Sigler, Mateusz Litwin, Scott Gray, Benjamin Chess, Jack Clark, Christopher Berner, Sam McCandlish, Alec Radford, Ilya Sutskever, and Dario Amodei.
\newblock Language {{Models}} are {{Few-Shot Learners}}.
\newblock In \emph{Advances in {{Neural Information Processing Systems}}}, volume~33, pp.\  1877--1901. Curran Associates, Inc., 2020.

\bibitem[Callaway(2024)]{callawayCouldAIdesignedProteins2024}
Ewen Callaway.
\newblock Could {{AI-designed}} proteins be weaponized? {{Scientists}} lay out safety guidelines.
\newblock \emph{Nature}, 627\penalty0 (8004):\penalty0 478--478, March 2024.
\newblock \doi{10.1038/d41586-024-00699-0}.

\bibitem[Carlini et~al.(2022)Carlini, Ippolito, Jagielski, Lee, Tramer, and Zhang]{carliniQuantifyingMemorizationNeural2022}
Nicholas Carlini, Daphne Ippolito, Matthew Jagielski, Katherine Lee, Florian Tramer, and Chiyuan Zhang.
\newblock Quantifying {{Memorization Across Neural Language Models}}.
\newblock In \emph{The {{Eleventh International Conference}} on {{Learning Representations}}}, September 2022.

\bibitem[Chen et~al.(2024)Chen, {Shwartz-Ziv}, Cho, Leavitt, and Saphra]{chenSuddenDropsLoss2024}
Angelica Chen, Ravid {Shwartz-Ziv}, Kyunghyun Cho, Matthew~L. Leavitt, and Naomi Saphra.
\newblock Sudden {{Drops}} in the {{Loss}}: {{Syntax Acquisition}}, {{Phase Transitions}}, and {{Simplicity Bias}} in {{MLMs}}, February 2024.

\bibitem[Chen et~al.(2021)Chen, Dao, Liang, Yang, Song, Rudra, and Re]{chenPixelatedButterflySimple2021}
Beidi Chen, Tri Dao, Kaizhao Liang, Jiaming Yang, Zhao Song, Atri Rudra, and Christopher Re.
\newblock Pixelated {{Butterfly}}: {{Simple}} and {{Efficient Sparse}} training for {{Neural Network Models}}.
\newblock In \emph{International {{Conference}} on {{Learning Representations}}}, October 2021.

\bibitem[Corbett et~al.(2012)Corbett, Dean, Epstein, Fikes, Frost, Furman, Ghemawat, Gubarev, Heiser, Hochschild, Hsieh, Kanthak, Kogan, Li, Lloyd, Melnik, Mwaura, Nagle, Quinlan, Rao, Rolig, Saito, Szymaniak, Taylor, Wang, and Woodford]{corbettSpannerGooglesGloballyDistributed2012}
James~C. Corbett, Jeffrey Dean, Michael Epstein, Andrew Fikes, Christopher Frost, J.~J. Furman, Sanjay Ghemawat, Andrey Gubarev, Christopher Heiser, Peter Hochschild, Wilson Hsieh, Sebastian Kanthak, Eugene Kogan, Hongyi Li, Alexander Lloyd, Sergey Melnik, David Mwaura, David Nagle, Sean Quinlan, Rajesh Rao, Lindsay Rolig, Yasushi Saito, Michal Szymaniak, Christopher Taylor, Ruth Wang, and Dale Woodford.
\newblock Spanner: {{Google}}'s {{Globally-Distributed Database}}.
\newblock In \emph{10th {{USENIX Symposium}} on {{Operating Systems Design}} and {{Implementation}} ({{OSDI}} 12)}, pp.\  261--264, 2012.
\newblock ISBN 978-1-931971-96-6.

\bibitem[Dafoe(2018)]{dafoeAIGovernanceResearch2018}
Allan Dafoe.
\newblock {{AI Governance}}: {{A Research Agenda}}, August 2018.

\bibitem[Dao \& Gu(2024)Dao and Gu]{daoTransformersAreSSMs2024}
Tri Dao and Albert Gu.
\newblock Transformers are {{SSMs}}: {{Generalized Models}} and {{Efficient Algorithms Through Structured State Space Duality}}.
\newblock https://arxiv.org/abs/2405.21060v1, May 2024.

\bibitem[Dean \& Ghemawat(2008)Dean and Ghemawat]{deanMapReduceSimplifiedData2008}
Jeffrey Dean and Sanjay Ghemawat.
\newblock {{MapReduce}}: Simplified data processing on large clusters.
\newblock \emph{Commun. ACM}, 51\penalty0 (1):\penalty0 107--113, January 2008.
\newblock ISSN 0001-0782.
\newblock \doi{10.1145/1327452.1327492}.

\bibitem[Deng et~al.(2023)Deng, Wang, Feng, Deng, Wang, and He]{dengAttackPromptGeneration2023}
Boyi Deng, Wenjie Wang, Fuli Feng, Yang Deng, Qifan Wang, and Xiangnan He.
\newblock Attack {{Prompt Generation}} for {{Red Teaming}} and {{Defending Large Language Models}}.
\newblock In Houda Bouamor, Juan Pino, and Kalika Bali (eds.), \emph{Findings of the {{Association}} for {{Computational Linguistics}}: {{EMNLP}} 2023}, pp.\  2176--2189, Singapore, December 2023. Association for Computational Linguistics.
\newblock \doi{10.18653/v1/2023.findings-emnlp.143}.

\bibitem[Deng et~al.(2009)Deng, Dong, Socher, Li, Li, and Li]{dengImageNetLargescaleHierarchical2009}
Jia Deng, Wei Dong, Richard Socher, Li-Jia Li, Kai Li, and Fei-Fei Li.
\newblock {{ImageNet}}: {{A}} large-scale hierarchical image database.
\newblock In \emph{2009 {{IEEE Conference}} on {{Computer Vision}} and {{Pattern Recognition}}\vphantom\{\}}, pp.\  248--255, 2009.
\newblock \doi{10.1109/CVPR.2009.5206848}.

\bibitem[Dragan et~al.(2024)Dragan, King, and Dafoe]{draganFrontierSafetyFramework2024}
Anca Dragan, Helen King, and Allan Dafoe.
\newblock Frontier {{Safety Framework}}, May 2024.

\bibitem[Eldan \& Li(2023)Eldan and Li]{eldanTinyStoriesHowSmall2023}
Ronen Eldan and Yuanzhi Li.
\newblock {{TinyStories}}: {{How Small Can Language Models Be}} and {{Still Speak Coherent English}}?, May 2023.

\bibitem[{European Union}(2024)]{europeanunionRegulationEU20242024}
{European Union}.
\newblock Regulation ({{EU}}) 2024/1689 of the {{European Parliament}} and of the {{Council}} of 13 {{June}} 2024 laying down harmonised rules on artificial intelligence and amending {{Regulations}} ({{EC}}) {{No}} 300/2008, ({{EU}}) {{No}} 167/2013, ({{EU}}) {{No}} 168/2013, ({{EU}}) 2018/858, ({{EU}}) 2018/1139 and ({{EU}}) 2019/2144 and {{Directives}} 2014/90/{{EU}}, ({{EU}}) 2016/797 and ({{EU}}) 2020/1828 ({{Artificial Intelligence Act}}), July 2024.

\bibitem[Feldman(2020)]{feldmanDoesLearningRequire2020}
Vitaly Feldman.
\newblock Does learning require memorization? a short tale about a long tail.
\newblock In \emph{Proceedings of the 52nd {{Annual ACM SIGACT Symposium}} on {{Theory}} of {{Computing}}}, {{STOC}} 2020, pp.\  954--959, New York, NY, USA, June 2020. Association for Computing Machinery.
\newblock ISBN 978-1-4503-6979-4.
\newblock \doi{10.1145/3357713.3384290}.

\bibitem[Feldman \& Zhang(2020)Feldman and Zhang]{feldmanWhatNeuralNetworks2020}
Vitaly Feldman and Chiyuan Zhang.
\newblock What {{Neural Networks Memorize}} and {{Why}}: {{Discovering}} the {{Long Tail}} via {{Influence Estimation}}.
\newblock In \emph{Advances in {{Neural Information Processing Systems}}}, volume~33, pp.\  2881--2891. Curran Associates, Inc., 2020.

\bibitem[Fleisig et~al.(2024)Fleisig, Smith, Bossi, Rustagi, Yin, and Klein]{fleisigLinguisticBiasChatGPT2024}
Eve Fleisig, Genevieve Smith, Madeline Bossi, Ishita Rustagi, Xavier Yin, and Dan Klein.
\newblock Linguistic {{Bias}} in {{ChatGPT}}: {{Language Models Reinforce Dialect Discrimination}}, September 2024.

\bibitem[Ganguli et~al.(2022)Ganguli, Lovitt, Kernion, Askell, Bai, Kadavath, Mann, Perez, Schiefer, Ndousse, Jones, Bowman, Chen, Conerly, DasSarma, Drain, Elhage, {El-Showk}, Fort, {Hatfield-Dodds}, Henighan, Hernandez, Hume, Jacobson, Johnston, Kravec, Olsson, Ringer, {Tran-Johnson}, Amodei, Brown, Joseph, McCandlish, Olah, Kaplan, and Clark]{ganguliRedTeamingLanguage2022}
Deep Ganguli, Liane Lovitt, Jackson Kernion, Amanda Askell, Yuntao Bai, Saurav Kadavath, Ben Mann, Ethan Perez, Nicholas Schiefer, Kamal Ndousse, Andy Jones, Sam Bowman, Anna Chen, Tom Conerly, Nova DasSarma, Dawn Drain, Nelson Elhage, Sheer {El-Showk}, Stanislav Fort, Zac {Hatfield-Dodds}, Tom Henighan, Danny Hernandez, Tristan Hume, Josh Jacobson, Scott Johnston, Shauna Kravec, Catherine Olsson, Sam Ringer, Eli {Tran-Johnson}, Dario Amodei, Tom Brown, Nicholas Joseph, Sam McCandlish, Chris Olah, Jared Kaplan, and Jack Clark.
\newblock Red {{Teaming Language Models}} to {{Reduce Harms}}: {{Methods}}, {{Scaling Behaviors}}, and {{Lessons Learned}}, November 2022.

\bibitem[Gao et~al.(2024)Gao, Xiong, Gao, Jia, Pan, Bi, Dai, Sun, Wang, and Wang]{gaoRetrievalAugmentedGenerationLarge2024}
Yunfan Gao, Yun Xiong, Xinyu Gao, Kangxiang Jia, Jinliu Pan, Yuxi Bi, Yi~Dai, Jiawei Sun, Meng Wang, and Haofen Wang.
\newblock Retrieval-{{Augmented Generation}} for {{Large Language Models}}: {{A Survey}}, March 2024.

\bibitem[Gebru et~al.(2021)Gebru, Morgenstern, Vecchione, Vaughan, Wallach, Daum{\'e}~III, and Crawford]{gebruDatasheetsDatasets2021}
Timnit Gebru, Jamie Morgenstern, Briana Vecchione, Jennifer~Wortman Vaughan, Hanna Wallach, Hal Daum{\'e}~III, and Kate Crawford.
\newblock Datasheets for {{Datasets}}, December 2021.

\bibitem[{Gemini Team}(2024{\natexlab{a}})]{geminiteamGemini15Unlocking2024}
{Gemini Team}.
\newblock Gemini 1.5: {{Unlocking}} multimodal understanding across millions of tokens of context, March 2024{\natexlab{a}}.

\bibitem[{Gemini Team}(2024{\natexlab{b}})]{geminiteamGeminiFamilyHighly2024}
{Gemini Team}.
\newblock Gemini: {{A Family}} of {{Highly Capable Multimodal Models}}, June 2024{\natexlab{b}}.

\bibitem[{Gemma Team}(2024)]{gemmateamGemmaOpenModels2024}
{Gemma Team}.
\newblock Gemma: {{Open Models Based}} on {{Gemini Research}} and {{Technology}}, March 2024.

\bibitem[{Google}(2023)]{googlePaLMTechnicalReport2023}
{Google}.
\newblock {{PaLM}} 2 {{Technical Report}}, May 2023.

\bibitem[{Google}(2024)]{googlePaliGemmaModelCard2024}
{Google}.
\newblock {{PaliGemma}} model card.
\newblock https://ai.google.dev/gemma/docs/paligemma/model-card, 2024.

\bibitem[Gupta et~al.(2019)Gupta, Hosfelt, Sajeev, Patel, Goodman, Doshi, Heim, Choset, and Gaston]{guptaXBDDatasetAssessing2019}
Ritwik Gupta, Richard Hosfelt, Sandra Sajeev, Nirav Patel, Bryce Goodman, Jigar Doshi, Eric Heim, Howie Choset, and Matthew Gaston.
\newblock {{xBD}}: {{A Dataset}} for {{Assessing Building Damage}} from {{Satellite Imagery}}, November 2019.

\bibitem[Gupta et~al.(2024)Gupta, Walker, Glickman, Koizumi, Bhatnagar, and Reddie]{guptaOpenSourceAssessmentsAI2024}
Ritwik Gupta, Leah Walker, Eli Glickman, Raine Koizumi, Sarthak Bhatnagar, and Andrew~W. Reddie.
\newblock Open-{{Source Assessments}} of {{AI Capabilities}}: {{The Proliferation}} of {{AI Analysis Tools}}, {{Replicating Competitor Models}}, and the {{Zhousidun Dataset}}, May 2024.

\bibitem[Hendrycks et~al.(2020)Hendrycks, Burns, Basart, Zou, Mazeika, Song, and Steinhardt]{hendrycksMeasuringMassiveMultitask2020}
Dan Hendrycks, Collin Burns, Steven Basart, Andy Zou, Mantas Mazeika, Dawn Song, and Jacob Steinhardt.
\newblock Measuring {{Massive Multitask Language Understanding}}.
\newblock In \emph{International {{Conference}} on {{Learning Representations}}}, October 2020.

\bibitem[Hilbert \& L{\'o}pez(2011)Hilbert and L{\'o}pez]{hilbertWorldsTechnologicalCapacity2011}
Martin Hilbert and Priscila L{\'o}pez.
\newblock The {{World}}'s {{Technological Capacity}} to {{Store}}, {{Communicate}}, and {{Compute Information}}.
\newblock \emph{Science}, 332\penalty0 (6025):\penalty0 60--65, April 2011.
\newblock \doi{10.1126/science.1200970}.

\bibitem[Hoffman \& Frase(2023)Hoffman and Frase]{hoffmanAddingStructureAI2023}
Mia Hoffman and Heather Frase.
\newblock Adding {{Structure}} to {{AI Harm}}: {{An Introduction}} to {{CSET}}'s {{AI Harm Framework}}.
\newblock Technical report, {Center for Security and Emerging Technology}, July 2023.

\bibitem[Hoffmann et~al.(2022{\natexlab{a}})Hoffmann, Borgeaud, Mensch, Buchatskaya, Cai, Rutherford, Casas, Hendricks, Welbl, Clark, Hennigan, Noland, Millican, van~den Driessche, Damoc, Guy, Osindero, Simonyan, Elsen, Rae, Vinyals, and Sifre]{hoffmannTrainingComputeOptimalLarge2022}
Jordan Hoffmann, Sebastian Borgeaud, Arthur Mensch, Elena Buchatskaya, Trevor Cai, Eliza Rutherford, Diego de~Las Casas, Lisa~Anne Hendricks, Johannes Welbl, Aidan Clark, Tom Hennigan, Eric Noland, Katie Millican, George van~den Driessche, Bogdan Damoc, Aurelia Guy, Simon Osindero, Karen Simonyan, Erich Elsen, Jack~W. Rae, Oriol Vinyals, and Laurent Sifre.
\newblock Training {{Compute-Optimal Large Language Models}}, March 2022{\natexlab{a}}.

\bibitem[Hoffmann et~al.(2022{\natexlab{b}})Hoffmann, Borgeaud, Mensch, Buchatskaya, Cai, Rutherford, {de Las Casas}, Hendricks, Welbl, Clark, Hennigan, Noland, Millican, {van den Driessche}, Damoc, Guy, Osindero, Simonyan, Elsen, Vinyals, Rae, and Sifre]{hoffmannEmpiricalAnalysisComputeoptimal2022}
Jordan Hoffmann, Sebastian Borgeaud, Arthur Mensch, Elena Buchatskaya, Trevor Cai, Eliza Rutherford, Diego {de Las Casas}, Lisa~Anne Hendricks, Johannes Welbl, Aidan Clark, Thomas Hennigan, Eric Noland, Katherine Millican, George {van den Driessche}, Bogdan Damoc, Aurelia Guy, Simon Osindero, Kar{\'e}n Simonyan, Erich Elsen, Oriol Vinyals, Jack Rae, and Laurent Sifre.
\newblock An empirical analysis of compute-optimal large language model training.
\newblock \emph{Advances in Neural Information Processing Systems}, 35:\penalty0 30016--30030, December 2022{\natexlab{b}}.

\bibitem[Huang et~al.(2023)Huang, Sun, Xie, Li, and Liu]{huangT2ICompBenchComprehensiveBenchmark2023}
Kaiyi Huang, Kaiyue Sun, Enze Xie, Zhenguo Li, and Xihui Liu.
\newblock {{T2I-CompBench}}: {{A Comprehensive Benchmark}} for {{Open-world Compositional Text-to-image Generation}}, October 2023.

\bibitem[Huh et~al.(2024)Huh, Cheung, Bernstein, Isola, and Agrawal]{huhTrainingNeuralNetworks2024}
Minyoung Huh, Brian Cheung, Jeremy Bernstein, Phillip Isola, and Pulkit Agrawal.
\newblock Training {{Neural Networks}} from {{Scratch}} with {{Parallel Low-Rank Adapters}}, July 2024.

\bibitem[Inan et~al.(2023)Inan, Upasani, Chi, Rungta, Iyer, Mao, Tontchev, Hu, Fuller, Testuggine, and Khabsa]{inanLlamaGuardLLMbased2023}
Hakan Inan, Kartikeya Upasani, Jianfeng Chi, Rashi Rungta, Krithika Iyer, Yuning Mao, Michael Tontchev, Qing Hu, Brian Fuller, Davide Testuggine, and Madian Khabsa.
\newblock Llama {{Guard}}: {{LLM-based Input-Output Safeguard}} for {{Human-AI Conversations}}.
\newblock https://arxiv.org/abs/2312.06674v1, December 2023.

\bibitem[Jiang et~al.(2023)Jiang, Sablayrolles, Mensch, Bamford, Chaplot, de~las Casas, Bressand, Lengyel, Lample, Saulnier, Lavaud, Lachaux, Stock, Scao, Lavril, Wang, Lacroix, and Sayed]{jiangMistral7B2023}
Albert~Q. Jiang, Alexandre Sablayrolles, Arthur Mensch, Chris Bamford, Devendra~Singh Chaplot, Diego de~las Casas, Florian Bressand, Gianna Lengyel, Guillaume Lample, Lucile Saulnier, L{\'e}lio~Renard Lavaud, Marie-Anne Lachaux, Pierre Stock, Teven~Le Scao, Thibaut Lavril, Thomas Wang, Timoth{\'e}e Lacroix, and William~El Sayed.
\newblock Mistral {{7B}}, October 2023.

\bibitem[Kaplan et~al.(2020)Kaplan, McCandlish, Henighan, Brown, Chess, Child, Gray, Radford, Wu, and Amodei]{kaplanScalingLawsNeural2020}
Jared Kaplan, Sam McCandlish, Tom Henighan, Tom~B. Brown, Benjamin Chess, Rewon Child, Scott Gray, Alec Radford, Jeffrey Wu, and Dario Amodei.
\newblock Scaling {{Laws}} for {{Neural Language Models}}, January 2020.

\bibitem[Kazemzadeh et~al.(2014)Kazemzadeh, Ordonez, Matten, and Berg]{kazemzadehReferItGameReferringObjects2014}
Sahar Kazemzadeh, Vicente Ordonez, Mark Matten, and Tamara Berg.
\newblock {{ReferItGame}}: {{Referring}} to {{Objects}} in {{Photographs}} of {{Natural Scenes}}.
\newblock In Alessandro Moschitti, Bo~Pang, and Walter Daelemans (eds.), \emph{Proceedings of the 2014 {{Conference}} on {{Empirical Methods}} in {{Natural Language Processing}} ({{EMNLP}})}, pp.\  787--798, Doha, Qatar, October 2014. Association for Computational Linguistics.
\newblock \doi{10.3115/v1/D14-1086}.

\bibitem[Kong et~al.(2024)Kong, Goel, Badlani, Ping, Valle, and Catanzaro]{kongAudioFlamingoNovel2024}
Zhifeng Kong, Arushi Goel, Rohan Badlani, Wei Ping, Rafael Valle, and Bryan Catanzaro.
\newblock Audio {{Flamingo}}: {{A Novel Audio Language Model}} with {{Few-Shot Learning}} and {{Dialogue Abilities}}, May 2024.

\bibitem[Leiserson et~al.(2020)Leiserson, Thompson, Emer, Kuszmaul, Lampson, Sanchez, and Schardl]{leisersonTheresPlentyRoom2020}
Charles~E. Leiserson, Neil~C. Thompson, Joel~S. Emer, Bradley~C. Kuszmaul, Butler~W. Lampson, Daniel Sanchez, and Tao~B. Schardl.
\newblock There's plenty of room at the {{Top}}: {{What}} will drive computer performance after {{Moore}}'s law?
\newblock \emph{Science}, 368\penalty0 (6495):\penalty0 eaam9744, June 2020.
\newblock \doi{10.1126/science.aam9744}.

\bibitem[Lewis et~al.(2020)Lewis, Perez, Piktus, Petroni, Karpukhin, Goyal, K{\"u}ttler, Lewis, Yih, Rockt{\"a}schel, Riedel, and Kiela]{lewisRetrievalAugmentedGenerationKnowledgeIntensive2020}
Patrick Lewis, Ethan Perez, Aleksandra Piktus, Fabio Petroni, Vladimir Karpukhin, Naman Goyal, Heinrich K{\"u}ttler, Mike Lewis, Wen-tau Yih, Tim Rockt{\"a}schel, Sebastian Riedel, and Douwe Kiela.
\newblock Retrieval-{{Augmented Generation}} for {{Knowledge-Intensive NLP Tasks}}.
\newblock In \emph{Advances in {{Neural Information Processing Systems}}}, volume~33, pp.\  9459--9474. Curran Associates, Inc., 2020.

\bibitem[Liang et~al.(2023)Liang, Bommasani, Lee, Tsipras, Soylu, Yasunaga, Zhang, Narayanan, Wu, Kumar, Newman, Yuan, Yan, Zhang, Cosgrove, Manning, R{\'e}, {Acosta-Navas}, Hudson, Zelikman, Durmus, Ladhak, Rong, Ren, Yao, Wang, Santhanam, Orr, Zheng, Yuksekgonul, Suzgun, Kim, Guha, Chatterji, Khattab, Henderson, Huang, Chi, Xie, Santurkar, Ganguli, Hashimoto, Icard, Zhang, Chaudhary, Wang, Li, Mai, Zhang, and Koreeda]{liangHolisticEvaluationLanguage2023}
Percy Liang, Rishi Bommasani, Tony Lee, Dimitris Tsipras, Dilara Soylu, Michihiro Yasunaga, Yian Zhang, Deepak Narayanan, Yuhuai Wu, Ananya Kumar, Benjamin Newman, Binhang Yuan, Bobby Yan, Ce~Zhang, Christian Cosgrove, Christopher~D. Manning, Christopher R{\'e}, Diana {Acosta-Navas}, Drew~A. Hudson, Eric Zelikman, Esin Durmus, Faisal Ladhak, Frieda Rong, Hongyu Ren, Huaxiu Yao, Jue Wang, Keshav Santhanam, Laurel Orr, Lucia Zheng, Mert Yuksekgonul, Mirac Suzgun, Nathan Kim, Neel Guha, Niladri Chatterji, Omar Khattab, Peter Henderson, Qian Huang, Ryan Chi, Sang~Michael Xie, Shibani Santurkar, Surya Ganguli, Tatsunori Hashimoto, Thomas Icard, Tianyi Zhang, Vishrav Chaudhary, William Wang, Xuechen Li, Yifan Mai, Yuhui Zhang, and Yuta Koreeda.
\newblock Holistic {{Evaluation}} of {{Language Models}}, October 2023.

\bibitem[Liu et~al.(2024)Liu, Yan, Zaharia, and Abbeel]{liuWorldModelMillionLength2024}
Hao Liu, Wilson Yan, Matei Zaharia, and Pieter Abbeel.
\newblock World {{Model}} on {{Million-Length Video And Language With Blockwise RingAttention}}, July 2024.

\bibitem[Liu et~al.(2023)Liu, Zhang, Wang, Wang, Yang, and Tang]{liuUniversalSegmentationArbitrary2023}
Yong Liu, Cairong Zhang, Yitong Wang, Jiahao Wang, Yujiu Yang, and Yansong Tang.
\newblock Universal {{Segmentation}} at {{Arbitrary Granularity}} with {{Language Instruction}}, December 2023.

\bibitem[Longpre et~al.(2023)Longpre, Mahari, Chen, {Obeng-Marnu}, Sileo, Brannon, Muennighoff, Khazam, Kabbara, Perisetla, Wu, Shippole, Bollacker, Wu, Villa, Pentland, and Hooker]{longpreDataProvenanceInitiative2023}
Shayne Longpre, Robert Mahari, Anthony Chen, Naana {Obeng-Marnu}, Damien Sileo, William Brannon, Niklas Muennighoff, Nathan Khazam, Jad Kabbara, Kartik Perisetla, Xinyi Wu, Enrico Shippole, Kurt Bollacker, Tongshuang Wu, Luis Villa, Sandy Pentland, and Sara Hooker.
\newblock The {{Data Provenance Initiative}}: {{A Large Scale Audit}} of {{Dataset Licensing}} \& {{Attribution}} in {{AI}}, November 2023.

\bibitem[McGuinness(2023)]{mcguinnessGPT4DetailsRevealed2023}
Patrick McGuinness.
\newblock {{GPT-4 Details Revealed}}, July 2023.

\bibitem[Mitchell et~al.(2019)Mitchell, Wu, Zaldivar, Barnes, Vasserman, Hutchinson, Spitzer, Raji, and Gebru]{mitchellModelCardsModel2019}
Margaret Mitchell, Simone Wu, Andrew Zaldivar, Parker Barnes, Lucy Vasserman, Ben Hutchinson, Elena Spitzer, Inioluwa~Deborah Raji, and Timnit Gebru.
\newblock Model {{Cards}} for {{Model Reporting}}.
\newblock In \emph{Proceedings of the {{Conference}} on {{Fairness}}, {{Accountability}}, and {{Transparency}}}, pp.\  220--229, January 2019.
\newblock \doi{10.1145/3287560.3287596}.

\bibitem[Moore(1998)]{mooreCrammingMoreComponents1998}
Gordon~E Moore.
\newblock Cramming {{More Components}} onto {{Integrated Circuits}}.
\newblock \emph{PROCEEDINGS OF THE IEEE}, 86\penalty0 (1), 1998.

\bibitem[Mora et~al.(2024)Mora, Wong, Lepe, Bhatia, Elmaaroufi, Varghese, Gonzalez, Polgreen, and Seshia]{moraSyntheticProgrammingElicitation2024}
Federico Mora, Justin Wong, Haley Lepe, Sahil Bhatia, Karim Elmaaroufi, George Varghese, Joseph~E. Gonzalez, Elizabeth Polgreen, and Sanjit~A. Seshia.
\newblock Synthetic {{Programming Elicitation}} and {{Repair}} for {{Text-to-Code}} in {{Very Low-Resource Programming Languages}}, June 2024.

\bibitem[Nasr et~al.(2023)Nasr, Carlini, Hayase, Jagielski, Cooper, Ippolito, {Choquette-Choo}, Wallace, Tram{\`e}r, and Lee]{nasrScalableExtractionTraining2023}
Milad Nasr, Nicholas Carlini, Jonathan Hayase, Matthew Jagielski, A.~Feder Cooper, Daphne Ippolito, Christopher~A. {Choquette-Choo}, Eric Wallace, Florian Tram{\`e}r, and Katherine Lee.
\newblock Scalable {{Extraction}} of {{Training Data}} from ({{Production}}) {{Language Models}}, November 2023.

\bibitem[Ng et~al.(2021)Ng, Laird, and He]{ngDataCentricAICompetition2021}
Andrew Ng, Dillon Laird, and Lynn He.
\newblock Data-{{Centric AI Competition}}, 2021.

\bibitem[{OpenAI}(2024{\natexlab{a}})]{openaiDisruptingDeceptiveUses2024}
{OpenAI}.
\newblock Disrupting deceptive uses of {{AI}} by covert influence operations.
\newblock https://openai.com/index/disrupting-deceptive-uses-of-AI-by-covert-influence-operations/, May 2024{\natexlab{a}}.

\bibitem[{OpenAI}(2024{\natexlab{b}})]{openaiDisruptingMaliciousUses2024}
{OpenAI}.
\newblock Disrupting malicious uses of {{AI}} by state-affiliated threat actors.
\newblock https://openai.com/index/disrupting-malicious-uses-of-ai-by-state-affiliated-threat-actors/, February 2024{\natexlab{b}}.

\bibitem[{OpenAI}(2024{\natexlab{c}})]{openaiGPT4TechnicalReport2024}
{OpenAI}.
\newblock {{GPT-4 Technical Report}}, March 2024{\natexlab{c}}.

\bibitem[Ouyang et~al.(2022)Ouyang, Wu, Jiang, Almeida, Wainwright, Mishkin, Zhang, Agarwal, Slama, Ray, Schulman, Hilton, Kelton, Miller, Simens, Askell, Welinder, Christiano, Leike, and Lowe]{ouyangTrainingLanguageModels2022}
Long Ouyang, Jeffrey Wu, Xu~Jiang, Diogo Almeida, Carroll Wainwright, Pamela Mishkin, Chong Zhang, Sandhini Agarwal, Katarina Slama, Alex Ray, John Schulman, Jacob Hilton, Fraser Kelton, Luke Miller, Maddie Simens, Amanda Askell, Peter Welinder, Paul~F. Christiano, Jan Leike, and Ryan Lowe.
\newblock Training language models to follow instructions with human feedback.
\newblock \emph{Advances in Neural Information Processing Systems}, 35:\penalty0 27730--27744, December 2022.

\bibitem[Perez et~al.(2022)Perez, Huang, Song, Cai, Ring, Aslanides, Glaese, McAleese, and Irving]{perezRedTeamingLanguage2022}
Ethan Perez, Saffron Huang, Francis Song, Trevor Cai, Roman Ring, John Aslanides, Amelia Glaese, Nat McAleese, and Geoffrey Irving.
\newblock Red {{Teaming Language Models}} with {{Language Models}}.
\newblock In Yoav Goldberg, Zornitsa Kozareva, and Yue Zhang (eds.), \emph{Proceedings of the 2022 {{Conference}} on {{Empirical Methods}} in {{Natural Language Processing}}}, pp.\  3419--3448, Abu Dhabi, United Arab Emirates, December 2022. Association for Computational Linguistics.
\newblock \doi{10.18653/v1/2022.emnlp-main.225}.

\bibitem[Pushkarna et~al.(2022)Pushkarna, Zaldivar, and Kjartansson]{pushkarnaDataCardsPurposeful2022}
Mahima Pushkarna, Andrew Zaldivar, and Oddur Kjartansson.
\newblock Data {{Cards}}: {{Purposeful}} and {{Transparent Dataset Documentation}} for {{Responsible AI}}.
\newblock In \emph{2022 {{ACM Conference}} on {{Fairness}}, {{Accountability}}, and {{Transparency}}}, pp.\  1776--1826, Seoul Republic of Korea, June 2022. ACM.
\newblock ISBN 978-1-4503-9352-2.
\newblock \doi{10.1145/3531146.3533231}.

\bibitem[Rahman et~al.(2024)Rahman, Owen, and You]{rahmanTrackingLargeScaleAI2024}
Robi Rahman, David Owen, and Josh You.
\newblock Tracking {{Large-Scale AI Models}}.
\newblock https://epochai.org/blog/tracking-large-scale-ai-models, April 2024.

\bibitem[Raji et~al.(2021)Raji, Bender, Paullada, Denton, and Hanna]{rajiAIEverythingWhole2021}
Inioluwa~Deborah Raji, Emily~M. Bender, Amandalynne Paullada, Emily Denton, and Alex Hanna.
\newblock {{AI}} and the {{Everything}} in the {{Whole Wide World Benchmark}}, November 2021.

\bibitem[Romney et~al.(2024)Romney, Reed, Moran, and King]{romneyAILetter2024}
Mitt Romney, Jack Reed, Jerry Moran, and Angus King.
\newblock {{AI Letter}}, April 2024.

\bibitem[Ruan et~al.(2024)Ruan, Maddison, and Hashimoto]{ruanObservationalScalingLaws2024}
Yangjun Ruan, Chris~J. Maddison, and Tatsunori Hashimoto.
\newblock Observational {{Scaling Laws}} and the {{Predictability}} of {{Language Model Performance}}.
\newblock https://arxiv.org/abs/2405.10938v2, May 2024.

\bibitem[Russakovsky et~al.(2014)Russakovsky, Deng, Su, Krause, Satheesh, Ma, Huang, Karpathy, Khosla, Bernstein, Berg, and {Fei-Fei}]{russakovskyImageNetLargeScale2014}
Olga Russakovsky, Jia Deng, Hao Su, Jonathan Krause, Sanjeev Satheesh, Sean Ma, Zhiheng Huang, Andrej Karpathy, Aditya Khosla, Michael Bernstein, Alexander~C. Berg, and Li~{Fei-Fei}.
\newblock {{ImageNet Large Scale Visual Recognition Challenge}}.
\newblock https://arxiv.org/abs/1409.0575v3, September 2014.

\bibitem[Schick et~al.(2023)Schick, {Dwivedi-Yu}, Dessi, Raileanu, Lomeli, Hambro, Zettlemoyer, Cancedda, and Scialom]{schickToolformerLanguageModels2023}
Timo Schick, Jane {Dwivedi-Yu}, Roberto Dessi, Roberta Raileanu, Maria Lomeli, Eric Hambro, Luke Zettlemoyer, Nicola Cancedda, and Thomas Scialom.
\newblock Toolformer: {{Language Models Can Teach Themselves}} to {{Use Tools}}.
\newblock In \emph{Thirty-Seventh {{Conference}} on {{Neural Information Processing Systems}}}, November 2023.

\bibitem[Schramowski et~al.(2022)Schramowski, Tauchmann, and Kersting]{schramowskiCanMachinesHelp2022}
Patrick Schramowski, Christopher Tauchmann, and Kristian Kersting.
\newblock Can {{Machines Help Us Answering Question}} 16 in {{Datasheets}}, and {{In Turn Reflecting}} on {{Inappropriate Content}}?
\newblock In \emph{Proceedings of the 2022 {{ACM Conference}} on {{Fairness}}, {{Accountability}}, and {{Transparency}}}, {{FAccT}} '22, pp.\  1350--1361, New York, NY, USA, June 2022. Association for Computing Machinery.
\newblock ISBN 978-1-4503-9352-2.
\newblock \doi{10.1145/3531146.3533192}.

\bibitem[Schuster \& Nakajima(2012)Schuster and Nakajima]{schusterJapaneseKoreanVoice2012}
Mike Schuster and Kaisuke Nakajima.
\newblock Japanese and {{Korean}} voice search.
\newblock In \emph{2012 {{IEEE International Conference}} on {{Acoustics}}, {{Speech}} and {{Signal Processing}} ({{ICASSP}})}, pp.\  5149--5152, March 2012.
\newblock \doi{10.1109/ICASSP.2012.6289079}.

\bibitem[Sennrich et~al.(2016)Sennrich, Haddow, and Birch]{sennrichNeuralMachineTranslation2016}
Rico Sennrich, Barry Haddow, and Alexandra Birch.
\newblock Neural {{Machine Translation}} of {{Rare Words}} with {{Subword Units}}.
\newblock In Katrin Erk and Noah~A. Smith (eds.), \emph{Proceedings of the 54th {{Annual Meeting}} of the {{Association}} for {{Computational Linguistics}} ({{Volume}} 1: {{Long Papers}})}, pp.\  1715--1725, Berlin, Germany, August 2016. Association for Computational Linguistics.
\newblock \doi{10.18653/v1/P16-1162}.

\bibitem[Shvachko et~al.(2010)Shvachko, Kuang, Radia, and Chansler]{shvachkoHadoopDistributedFile2010}
Konstantin Shvachko, Hairong Kuang, Sanjay Radia, and Robert Chansler.
\newblock The {{Hadoop Distributed File System}}.
\newblock In \emph{2010 {{IEEE}} 26th {{Symposium}} on {{Mass Storage Systems}} and {{Technologies}} ({{MSST}})}, pp.\  1--10, Incline Village, NV, USA, May 2010. IEEE.
\newblock ISBN 978-1-4244-7152-2.
\newblock \doi{10.1109/MSST.2010.5496972}.

\bibitem[Si et~al.(2024)Si, Yang, and Hashimoto]{siCanLLMsGenerate2024}
Chenglei Si, Diyi Yang, and Tatsunori Hashimoto.
\newblock Can {{LLMs Generate Novel Research Ideas}}? {{A Large-Scale Human Study}} with 100+ {{NLP Researchers}}, September 2024.

\bibitem[{The White House}(2023)]{thewhitehouseExecutiveOrderSafe2023}
{The White House}.
\newblock Executive {{Order}} on the {{Safe}}, {{Secure}}, and {{Trustworthy Development}} and {{Use}} of {{Artificial Intelligence}}.
\newblock https://www.whitehouse.gov/briefing-room/presidential-actions/2023/10/30/executive-order-on-the-safe-secure-and-trustworthy-development-and-use-of-artificial-intelligence/, October 2023.

\bibitem[Thiel(2023)]{thielIdentifyingEliminatingCSAM2023}
David Thiel.
\newblock Identifying and {{Eliminating CSAM}} in {{Generative ML Training Data}} and {{Models}}.
\newblock \emph{Stanford Digital Repository}, 2023.
\newblock \doi{10.25740/kh752sm9123}.

\bibitem[Touvron et~al.(2023)Touvron, Lavril, Izacard, Martinet, Lachaux, Lacroix, Rozi{\`e}re, Goyal, Hambro, Azhar, Rodriguez, Joulin, Grave, and Lample]{touvronLLaMAOpenEfficient2023}
Hugo Touvron, Thibaut Lavril, Gautier Izacard, Xavier Martinet, Marie-Anne Lachaux, Timoth{\'e}e Lacroix, Baptiste Rozi{\`e}re, Naman Goyal, Eric Hambro, Faisal Azhar, Aurelien Rodriguez, Armand Joulin, Edouard Grave, and Guillaume Lample.
\newblock {{LLaMA}}: {{Open}} and {{Efficient Foundation Language Models}}, February 2023.

\bibitem[Trinh et~al.(2024)Trinh, Wu, Le, He, and Luong]{trinhSolvingOlympiadGeometry2024}
Trieu~H. Trinh, Yuhuai Wu, Quoc~V. Le, He~He, and Thang Luong.
\newblock Solving olympiad geometry without human demonstrations.
\newblock \emph{Nature}, 625\penalty0 (7995):\penalty0 476--482, January 2024.
\newblock ISSN 1476-4687.
\newblock \doi{10.1038/s41586-023-06747-5}.

\bibitem[Van~Drie(2007)]{vandrieComputeraidedDrugDesign2007}
John~H. Van~Drie.
\newblock Computer-aided drug design: The next 20 years.
\newblock \emph{Journal of Computer-Aided Molecular Design}, 21\penalty0 (10-11):\penalty0 591--601, 2007.
\newblock ISSN 0920-654X.
\newblock \doi{10.1007/s10822-007-9142-y}.

\bibitem[Wang et~al.(2020)Wang, Zhu, Torralba, and Efros]{wangDatasetDistillation2020}
Tongzhou Wang, Jun-Yan Zhu, Antonio Torralba, and Alexei~A. Efros.
\newblock Dataset {{Distillation}}, February 2020.

\bibitem[Wang et~al.(2023)Wang, Li, Kallidromitis, Kato, Kozuka, and Darrell]{wangHierarchicalOpenvocabularyUniversal2023}
Xudong Wang, Shufan Li, Konstantinos Kallidromitis, Yusuke Kato, Kazuki Kozuka, and Trevor Darrell.
\newblock Hierarchical {{Open-vocabulary Universal Image Segmentation}}, December 2023.

\bibitem[Wei et~al.(2023)Wei, Haghtalab, and Steinhardt]{weiJailbrokenHowDoes2023}
Alexander Wei, Nika Haghtalab, and Jacob Steinhardt.
\newblock Jailbroken: {{How Does LLM Safety Training Fail}}?, July 2023.

\bibitem[Wei et~al.(2022)Wei, Tay, Bommasani, Raffel, Zoph, Borgeaud, Yogatama, Bosma, Zhou, Metzler, Chi, Hashimoto, Vinyals, Liang, Dean, and Fedus]{weiEmergentAbilitiesLarge2022}
Jason Wei, Yi~Tay, Rishi Bommasani, Colin Raffel, Barret Zoph, Sebastian Borgeaud, Dani Yogatama, Maarten Bosma, Denny Zhou, Donald Metzler, Ed~H. Chi, Tatsunori Hashimoto, Oriol Vinyals, Percy Liang, Jeff Dean, and William Fedus.
\newblock Emergent {{Abilities}} of {{Large Language Models}}.
\newblock \emph{Transactions on Machine Learning Research}, June 2022.
\newblock ISSN 2835-8856.

\bibitem[Wiener et~al.(2024)Wiener, Roth, Rubio, and Stern]{wienerSB1047SafeSecure2024}
Scott Wiener, Richard Roth, Susan Rubio, and Henry Stern.
\newblock {{SB-1047 Safe}} and {{Secure Innovation}} for {{Frontier Artificial Intelligence Models Act}}, September 2024.

\bibitem[Yan et~al.(2023)Yan, Jiang, Wu, Wang, Luo, Yuan, and Lu]{yanUniversalInstancePerception2023}
Bin Yan, Yi~Jiang, Jiannan Wu, Dong Wang, Ping Luo, Zehuan Yuan, and Huchuan Lu.
\newblock Universal {{Instance Perception}} as {{Object Discovery}} and {{Retrieval}}, August 2023.

\bibitem[Yasunaga et~al.(2023)Yasunaga, Aghajanyan, Shi, James, Leskovec, Liang, Lewis, Zettlemoyer, and Yih]{yasunagaRetrievalAugmentedMultimodalLanguage2023}
Michihiro Yasunaga, Armen Aghajanyan, Weijia Shi, Rich James, Jure Leskovec, Percy Liang, Mike Lewis, Luke Zettlemoyer, and Wen-tau Yih.
\newblock Retrieval-{{Augmented Multimodal Language Modeling}}, June 2023.

\bibitem[Yu et~al.(2023)Yu, Yang, Pelrine, Godbout, and Rabbany]{yuOpenClosedSmall2023}
Hao Yu, Zachary Yang, Kellin Pelrine, Jean~Francois Godbout, and Reihaneh Rabbany.
\newblock Open, {{Closed}}, or {{Small Language Models}} for {{Text Classification}}?, August 2023.

\bibitem[Yuan et~al.(2024)Yuan, Li, Huang, Ye, and Sun]{yuanTinyGPTVEfficientMultimodal2024}
Zhengqing Yuan, Zhaoxu Li, Weiran Huang, Yanfang Ye, and Lichao Sun.
\newblock {{TinyGPT-V}}: {{Efficient Multimodal Large Language Model}} via {{Small Backbones}}, June 2024.

\bibitem[Zeng et~al.(2024)Zeng, Klyman, Zhou, Yang, Pan, Jia, Song, Liang, and Li]{zengAIRiskCategorization2024}
Yi~Zeng, Kevin Klyman, Andy Zhou, Yu~Yang, Minzhou Pan, Ruoxi Jia, Dawn Song, Percy Liang, and Bo~Li.
\newblock {{AI Risk Categorization Decoded}} ({{AIR}} 2024): {{From Government Regulations}} to {{Corporate Policies}}, June 2024.

\bibitem[Zhai et~al.(2022)Zhai, Kolesnikov, Houlsby, and Beyer]{zhaiScalingVisionTransformers2022}
Xiaohua Zhai, Alexander Kolesnikov, Neil Houlsby, and Lucas Beyer.
\newblock Scaling {{Vision Transformers}}.
\newblock In \emph{Proceedings of the {{IEEE}}/{{CVF Conference}} on {{Computer Vision}} and {{Pattern Recognition}}}, pp.\  12104--12113, 2022.

\bibitem[Zhang et~al.(2024)Zhang, Patil, Jain, Shen, Zaharia, Stoica, and Gonzalez]{zhangRAFTAdaptingLanguage2024}
Tianjun Zhang, Shishir~G. Patil, Naman Jain, Sheng Shen, Matei Zaharia, Ion Stoica, and Joseph~E. Gonzalez.
\newblock {{RAFT}}: {{Adapting Language Model}} to {{Domain Specific RAG}}, March 2024.

\bibitem[Zhao et~al.(2021)Zhao, Mopuri, and Bilen]{zhaoDatasetCondensationGradient2021}
Bo~Zhao, Konda~Reddy Mopuri, and Hakan Bilen.
\newblock Dataset {{Condensation}} with {{Gradient Matching}}, March 2021.

\bibitem[Zhu et~al.(2024)Zhu, Zhang, Sifferman, Sheaves, Wang, Richmond, Zhou, and Eshraghian]{zhuScalableMatMulfreeLanguage2024}
Rui-Jie Zhu, Yu~Zhang, Ethan Sifferman, Tyler Sheaves, Yiqiao Wang, Dustin Richmond, Peng Zhou, and Jason~K. Eshraghian.
\newblock Scalable {{MatMul-free Language Modeling}}, June 2024.

\end{thebibliography}
\bibliographystyle{iclr2025_conference}

\end{document}